\documentclass[mnsc,nonblindrev]{informs3noheader}

\usepackage{endnotes}
\let\footnote=\endnote

\usepackage{natbib}
 \bibpunct[, ]{(}{)}{,}{a}{}{,}%
 \def\bibfont{\small}%
 \def\bibsep{\smallskipamount}%
 \def\bibhang{24pt}%

\TheoremsNumberedThrough     

\EquationsNumberedThrough    

\usepackage{rotating,subcaption}
\usepackage[margin=1in]{geometry}
\usepackage[utf8]{inputenc}
\usepackage{amsfonts,verbatim,multirow,placeins,cancel}
\usepackage{dsfont,etex}
\usepackage{amsmath,amssymb,float,xcolor,multicol,bbm}
\usepackage{graphicx}

\newtheorem{obs}{Observation}

\newcommand{\alg}{\textsc{ALG}}
\newcommand{\opt}{{\mathbb OPT}}
\newcommand{\firstfit}{\textsc{First-Fit}}

\newcommand{\localsearch}{\textsc{Local-Search}}
\newcommand{\cost}{{\mathbb Cost}}

\begin{document}

\RUNAUTHOR{Cohen, Keller, Mirrokni and Zadimoghaddam}

\RUNTITLE{Overcommitment in Cloud Services - Bin packing with Chance Constraints}

\TITLE{Overcommitment in Cloud Services -- Bin packing with Chance Constraints}

\ARTICLEAUTHORS{%
\AUTHOR{Maxime C. Cohen}
\AFF{Google Research, New York, NY 10011, \EMAIL{maxccohen@google.com}}
\AUTHOR{Philipp W. Keller}
\AFF{Google, Mountain View, CA 94043, \EMAIL{pkeller@google.com}}
\AUTHOR{Vahab Mirrokni}
\AFF{Google Research, New York, NY 10011, \EMAIL{mirrokni@google.com}}
\AUTHOR{Morteza Zadimoghaddam}
\AFF{Google Research, New York, NY 10011, \EMAIL{zadim@google.com}}
}

\ABSTRACT{
This paper considers a traditional problem of resource allocation, scheduling jobs on
machines. One such recent application is cloud computing, where jobs arrive in an online fashion with capacity requirements and need to be
immediately scheduled on physical machines in data centers.
It is often observed that the requested capacities are not fully utilized, hence offering an opportunity to employ an \emph{overcommitment policy}, i.e., selling resources beyond capacity. Setting the right overcommitment level can induce a significant cost reduction for the cloud provider, while only inducing a very low risk of violating capacity constraints. We introduce and study a model that quantifies the value of overcommitment by modeling the problem as a bin packing with chance constraints. We then propose an alternative formulation that transforms each chance constraint into a submodular function. We show that our model captures the risk pooling effect and can guide scheduling and overcommitment decisions. We also develop a family of online algorithms that are intuitive, easy to implement and provide a constant factor guarantee from optimal. Finally, we calibrate our model using realistic workload data, and test our approach in a practical setting. Our analysis and experiments illustrate the benefit of overcommitment in cloud services, and suggest a cost reduction of 1.5\% to 17\% depending on the provider's risk tolerance.
}%

\KEYWORDS{Bin packing, Approximation algorithms, Cloud computing, Overcommitment}   

\maketitle

\section{Introduction}\label{intro}

Bin packing is an important problem with numerous applications such as hospitals, call centers, filling up containers, loading trucks with weight capacity constraints, creating file backups and more recently, cloud computing. A cloud provider needs to decide how many physical machines to purchase in order to accommodate the incoming jobs efficiently. This is typically modeled as a bin packing optimization problem, where one minimizes the cost of acquiring the physical machines subject to a capacity constraint for each machine. The jobs are assumed to arrive in an online fashion according to some vaguely specified arrival process. In addition, the jobs come with a specific requirement, but the effective job size and duration are not exactly known until after the actual scheduling has occurred. In practice, job size and duration can be estimated from historical data. One straightforward way to schedule jobs is to assume that each job will fully utilize its requirement (e.g., if a job requests 32 CPU cores, the cloud provider allocates this exact amount for the job). However, there is empirical evidence, that most of the virtual machines do not use the full requested capacity. This offers an opportunity for the cloud provider to employ an  \emph{overcommitment policy}, i.e., to schedule sets of jobs with the total requirement exceeding the respective capacities of physical machines. On one hand, the provider faces the risk that usage exceeds the physical capacity, which can result in severe penalties (e.g., acquiring or reallocating machines on the fly, canceling and rescheduling running jobs, mitigating interventions, etc.). On the other hand, if many jobs do not fully utilize their requested resources, the provider can potentially reduce the costs significantly. This becomes even more impactful in the cloud computing market, which has become increasingly competitive in recent years as Google, Amazon, and Microsoft aim to replace private data centers. ``The race to zero price'' is a commonly used term for this industry, where cloud providers have cut their prices very aggressively. According to an article in Business Insider in January 2015: ``Amazon Web Services (AWS), for example, has cut its price 44 times during 2009-2015, while Microsoft and Google have both decreased prices multiple times to keep up with AWS''. In January 2015, RBC Capital's Mark Mahaney published a chart that perfectly captures this trend and shows that the average monthly cost per gigabyte of RAM, for a set of various workloads, has dropped significantly: AWS dropped prices 8\% from Oct. 2013 to Dec. 2014, while both Google and Microsoft cut prices by 6\% and 5\%, respectively, in the same period. Other companies who charge more, like Rackspace and AT\&T, dropped prices even more significantly.

As a result, designing the right overcommitment policy for servers has a clear potential to increase the cloud provider profit. The goal of this paper is to study this question, and propose a model that helps guiding this type of decisions. In particular, we explicitly model job size uncertainty to motivate new algorithms, and evaluate them on realistic workloads. 

Our model and approaches are not limited to cloud computing and can be applied to several resource allocation problems. However, we will illustrate most of the discussions and applications using examples borrowed from the cloud computing world. Note that describing the cloud infrastructure and hardware is beyond the scope of this paper. For surveys on cloud computing, see, for example \cite{dinh2013survey} and \cite{fox2009above}.

We propose to model the problem as a bin packing with chance constraints, i.e., the total load assigned to each machine should be below physical capacity with a high pre-specified probability. Chance constraints are a commonly used modeling tool to capture risks and constraints on random variables (\cite{charnes1963deterministic}). Introducing chance constraints to several continuous optimization problems was extensively studied in the literature (see, e.g., \cite{calafiore2006distributionally} and \cite{delage2010distributionally}). This paper is the first to incorporate capacity chance constraints in the bin packing problem, and to propose efficient algorithms to solve the problem. Using some results from distributionally robust optimization (\cite{calafiore2006distributionally}), we reformulate the problem as a bin packing with submodular capacity constraints. Our reformulations are exact under the assumption of independent Gaussian resource usages for the jobs. More generally, they provide an upper bound and a good practical approximation in the realistic case where the jobs' usages are arbitrarily distributed but bounded. 

Using some machinery from previous work (see \cite{goemans2009approximating}, and \cite{SvitkinaFleischer}), we show that for the bin packing problem with general monotone submodular constraints, it is impossible to find a solution within any reasonable factor from optimal (more precisely, $\frac{\sqrt{N}}{\ln(N)}$, where $N$ is the number of jobs). In this paper, we show that our problem can be solved using a class of simple online algorithms that guarantee a constant factor of 8/3 from optimal (Theorem \ref{thm:lazyapp}). This class of algorithms includes the commonly used \emph{Best-Fit} and \emph{First-Fit} heuristics. We also develop an improved constant guarantee of 9/4 for the online problem (Theorem \ref{firstfit94}), and a 2-approximation for the offline version (Theorem \ref{offline2}). We further refine our results to the case where a large number of jobs can be scheduled on each machine (i.e., each job has a small size relative to the machine capacity). In this regime, our approach asymptotically converges to a 4/3 approximation. More importantly, our model and algorithms allow us to draw interesting insights on how one should schedule jobs. In particular, our approach (i) translates to a transparent recipe on how to assign jobs to machines; (ii) explicitly exploits the risk pooling effect; and (iii) can be used to guide an overcommitment strategy that significantly reduces the cost of purchasing machines.

We apply our algorithm to a synthetic but realistic workload inspired by historical production workloads in Google data centers, and show that it yields good performance. In particular, our method reduces the necessary number of physical machines, while limiting the risk borne by the provider. Our analysis also formalizes intuitions and provides insights regarding effective job scheduling strategies in practical settings.

\subsection{Contributions}

Scheduling jobs on machines can be modeled as a
bin packing problem. Jobs arrive online with some 
requirements, and the scheduler decides how many machines to purchase and how to schedule the jobs. Assuming random job
sizes and limited machine capacities, one can formulate the problem
as a 0/1 integer program. The objective is to minimize the number of machines
required, subject to constraints on the capacity of each machine.
In this paper, we model the capacity constraints
as chance constraints, and study the potential benefit of
overcommitment. The contributions of the paper can
be summarized as follows.

\begin {itemize}
\item {\textit{Formulating the overcommitment bin packing problem.}}  \\We present an optimization formulation for scheduling jobs on machines, while allowing the provider to overcommit. We first model the problem as \emph{Bin Packing with Chance Constraints} (BPCC). Then, we present an alternative \emph{Submodular Bin Packing} (SMBP) formulation that explicitly captures the risk pooling effect on each machine. We show that the SMBP is equivalent to the BPCC under common assumptions (independent Gaussian usage distributions), and that it is distributionally robust for usages with given means and diagonal covariance matrix). Perhaps most importantly from a practical perspective, the SMBP provides an upper bound and a good approximation under generic independent distributions over bounded intervals (see Proposition \ref{robustSMBP}). This last setting is most common in today's cloud data centers, where virtual machines are sold as fixed-size units.

\item {\textit{Developing simple algorithms that guarantee a constant factor approximation from optimal.}}  \\We show that our (SMBP) problem can be solved by well-known online algorithms such as \emph{First-Fit} and \emph{Best-Fit}, while guaranteeing a constant factor of 8/3 from optimal (Theorem \ref{thm:lazyapp}). We further refine this result in the case where a large number of jobs can be scheduled on each machine, and obtain a 4/3 approximation asymptotically (Corollary \ref{atleastKjobs}). We also develop an improved constant guarantee of 9/4 for the online problem using \emph{First-Fit} (Theorem \ref{firstfit94}), and a 2 approximation for the offline version (Theorem \ref{offline2}).  We then use our analysis to infer how one should assign jobs to machines, and show how to obtain a nearly optimal assignment (Theorem \ref{thm:full-homogeneous}). 
  
\item {\textit{Using our model to draw practical insights on the overcommitment policy.}}
  \\Our approach translates to a transparent and meaningful recipe on how to assign jobs to machines by clustering similar jobs in terms of statistical information. In addition, our approach explicitly captures the risk pooling effect: as we assign more jobs to a given machine, the ``safety buffer'' needed for each job decreases. Finally, our approach can be used to guide a practical overcommitment strategy, where one can significantly reduce the cost of purchasing machines by allowing a low risk of violating capacity constraints.

\item {\textit{Calibrating and applying our model to a practical setting.}}
  \\We use realistic workload data inspired by Google Compute Engine to calibrate our model and test our results
  in a practical setting. We observe that our proposed algorithm outperforms other natural scheduling schemes, and realizes a cost saving of 1.5\% to 17\% relative to the no-overcommitment policy.

\end{itemize}

\subsection{Literature review}\label{lit}

This paper is related to different streams of literature. 

In the optimization literature, the problem of scheduling jobs on virtual machines has been studied extensively, and the bin packing problem is a common formulation. 
Hundreds of papers study the bin packing problem including many of its variations, such as 2D packing (e.g., \cite{pisinger2005two}), linear packing, packing by weight, packing by cost, online bin packing, etc. The basic bin packing problem is NP-hard, and \cite{delorme2016exactcsp} provide a recent survey of exact approaches. However, several simple online algorithms are often used in practice for large-scale instances. A common variation is the problem where jobs arrive online with sizes sampled independently from a known discrete distribution with integer support and must be immediately packed onto machines upon arrival. The size of a job is known when it arrives, and the goal is to minimize the number of non-empty machines (or equivalently, minimize the \emph{waste}, defined as the total unused space). For this variation, the \emph{sum-of-squares} heuristic represents the state-of-the-art. It is almost distribution-agnostic, and nearly universally optimal for most distributions by achieving sublinear waste in the number of items seen (see, \cite{csirik2006sum}). In \cite{gupta2012online}, the authors propose two algorithms based on gradient descent on a suitably defined Lagrangian relaxations of the bin packing linear program that achieve additive $O(\sqrt{N})$ waste relative to the optimal policy. This line of work bounds the expected waste for general classes of job size distribution in an asymptotic sense.

Worst-case analysis of (finite, deterministic) bin packing solutions has received a lot of attention as well. For deterministic capacity constraints, several efficient algorithms have been proposed. They can be applied online, and admit approximation guarantees in both online and offline settings. The offline version of the problem can be solved using $(1+\epsilon) OPT + 1$ bins in linear time (\cite{de1981bin}).
A number of heuristics can solve large-scale instances efficiently while guaranteeing a constant factor cost relative to optimal. For a survey on approximation algorithms for bin packing, see for example \cite{coffman1996approximation}. Three such widely used heuristics are First-Fit (FF), Next-Fit (NF) and Best-Fit (BF) (see, e.g., \cite{bays1977comparison}, \cite{keller2012analysis} and \cite{kenyon1996best}). FF assigns the newly arrived job to the first machine that can accommodate it, and purchase a new machine only if none of the existing ones can fit the new job. NF is similar to FF but continues to assign jobs from the current machine without going back to previous machines. BF uses a similar strategy but seeks to fit the newly arrived job to the machine with the smallest remaining capacity. While one can easily show that these heuristics provide a 2-approximation guarantee, improved factors were also developed under special assumptions. \cite{dosa2013first} provide a tight upper bound for the FF strategy, showing that it never needs more than $1.7 OPT$ machines for any input. The offline version of the problem also received a lot of attention, and the Best-Fit-Decreasing (BFD) and First-Fit-Decreasing (FFD) strategies are among the simplest (and most popular) heuristics for solving it. They operate like BF and FF but first rank all the jobs in decreasing order of size. \cite{dosa2007tight} show that the tight bound of FFD is $11/9 OPT + 6/9$.

Our problem differs as our goal is to schedule jobs \emph{before} observing the realization of their size. In this case, stochastic bin packing models, where the job durations are modeled as random variables, are particularly relevant.  \cite{coffman1980stochastic} consider the problem and study the asymptotic and convergence properties of the Next-Fit online algorithm.
\cite{lueker1983bin} considers the case where the job durations are drawn uniformly from intervals of the form $[a,b]$, and derive a lower bound on the asymptotic expected number of bins used in an optimum packing. 
However, unlike this and other asymptotic results where the jobs' sizes are known when scheduling occurs, we are interested in computing a solution that is feasible with high probability before observing the actual sizes.
Our objective is to assign the jobs to as few machines as possible such that the set of jobs assigned to each machine satisfies the capacity constraint with some given probability (say $99\%$). In other words, we are solving a stochastic optimization problem, and studying/analyzing different simple heuristic solutions to achieve this goal. 
To make the difference with the worst case analysis clear, we note that the worst case analysis becomes a special case of our problem when the objective probability threshold is set to $100\%$ (instead of $99\%$, or any other number strictly less than 1). The whole point of our paper is to exploit the stochastic structure of the problem in order to reduce the scheduling costs via overcommitment. 


In this paper, we consider an auxiliary deterministic bin packing problem with a linear cost but non-linear modified capacity constraints. In \cite{anily1994worst}, the authors consider general cost structures with linear capacity constraints. More precisely, the cost of a machine is assumed to be a concave and monotone function of the number of jobs in the machine. They show that the Next-Fit Increasing heuristic provides a worst-case bound of no more than 1.75, and an asymptotic worst-case bound of 1.691. 

The motivation behind this paper is similar to the overbooking policy for airline companies and hotels. It is very common for airlines to overbook and accept additional reservations for seats on a flight beyond the aircraft's seating capacity\footnotemark[1]\footnotetext[1]{http://www.forbes.com/2009/04/16/airline-tickets-flights-lifestyle-travel-airlines-overbooked.html}. Airline companies (and hotels) employ an overbooking strategy for several reasons, including: (i) no-shows (several passengers are not showing up to their flight, and the airline can predict the no-show rate for each itinerary); (ii) increasing the profit by reducing lost opportunities; and (iii) segmenting passengers (charging a higher price as we get closer to the flight). Note that in the context of this paper, the same motivation of no-shows applies. However, the inter-temporal price discrimination is beyond the scope of our model.
Several academic papers in operations research have studied the overbooking problem within the last forty years (see, e.g., \cite{rothstein1971airline}, \cite{rothstein1985or}, \cite{weatherford1992taxonomy}, \cite{subramanian1999airline} and \cite{karaesmen2004overbooking}). The methodology is often based on solving a dynamic program incorporating some prediction of the no-show rate. In our problem, we face a large-scale bin packing problem that needs to be solved online. Rather than deciding how many passengers (jobs) to accept and at what price, cloud providers today usually aim to avoid declining any reasonable workloads at a fixed list price\footnotemark[2]\footnotetext[2]{The "spot instances" provided by Amazon and other heavily discounted reduced-availability services are notable exceptions.}.
  
This paper is also related to the robust optimization literature, and especially to distributionally robust optimization. The goal is to solve an optimization problem where the input parameter distribution belongs to a family of distributions that share some properties (e.g., all the distributions with the same mean and covariance matrix) and consider the worst-case within the given family (concrete examples are presented in Section \ref{altform}). Examples of such work include: \cite{ghaoui2003worst},  \cite{bertsimas2005optimal}, \cite{calafiore2006distributionally} and \cite{delage2010distributionally}. That work aims to convert linear or convex (continuous) optimization problems with a chance constraint into tractable formulations. Our paper shares a similar motivation but considers a problem with integer variables. To the best of our knowledge, this paper is the first to develop efficient algorithms with constant approximation guarantees for the bin packing problem with capacity chance constraints. 

Large-scale cluster management in general is an important area of computer systems research. \cite{google2015borg} provide a full, modern example of a production system. Among the work on scheduling jobs, \cite{google2011vmpacking} propose a model that also has a certain submodular structure due to the potential for sharing memory pages between virtual machines (in contrast to the risk-pooling effect modeled in this paper). Much experimental work seek to evaluate the real-world performance of bin packing heuristics that also account for factors such as adverse interactions between jobs scheduled together, and the presence of multiple contended resources (see for example \cite{microsoft2011validating} and \cite{microsoft2013performanceaware}). While modeling these aspects is likely to complement the resource savings achieved with the stochastic model we propose, these papers capture fundamentally different efficiency gains arising from technological improvements and idiosyncratic properties of certain types (or combinations) of resources. In this paper, we limit our attention to the benefit and practicality of machine over-commitment in the case where a single key resource is in short supply. This applies directly to multi-resource settings if, for example, the relatively high cost of one resource makes over-provisioning the others worthwhile, or if there is simply an imbalance between the relative supply and demand for the various resources making one of the resources scarce.
 
\textbf{Structure of the paper.} In Section \ref{model}, we present our model and assumptions. Then, we present the results and insights for special cases in Section \ref{analysis}. In Section \ref{constantapprox}, we consider the general case and develop a class of efficient approximation algorithms that guarantee a constant factor from optimal. In Section \ref{concav}, we exploit the structure of the problem in order to obtain a nearly optimal assignment  and to draw practical insights. In Sections \ref{extens} and \ref{comps}, we present extensions and computational experiments using realistic data respectively. Finally, our conclusions are reported in Section \ref{concl}. Most of the proofs of the Theorems and Propositions are relegated to the Appendix.

\section{Model}\label{model}

In this section, we present the model and assumptions we impose. We start by formulating the problem we want to solve, and then propose an alternative formulation. As we previously discussed, job requests for cloud services (or any other resource allocation problem) come with a requested capacity. This can be the memory or CPU requirements for virtual machines in the context of cloud computing, or job duration in more traditional scheduling problems where jobs are processed sequentially\footnotemark[3]\footnotetext[3]{Although there is also a job duration in cloud computing, it is generally unbounded and hence, even less constrained than the resource usage from the customer's perspective. The duration is also less important than the resource usage, since most virtual machines tend to be long-lived, cannot be delayed or pre-empted, and are paid for by the minute. In contrast, over-allocating unused, already paid-for resources can have a large impact on efficiency.}. We refer to $A_j$ as the size of job $j$ and assume that $A_j$ is a random variable. Historical data can provide insight into the distribution of $A_j$. For simplicity, we first consider the offline version of the problem where all the jobs arrive simultaneously at time 0, and our goal is to pack the jobs onto the minimum possible number of machines. Jobs cannot be delayed or preempted. The methods we develop in this paper can be applied in the more interesting online version of the problem, as we discuss in Section \ref{constantapprox}. We denote the capacity of
machine $i$ by $V_{i}$. Motivated by practical problems, and in accordance with prior work, we assume that all the machines have the same capacity, i.e., $V_i = V; \ \forall i$. In addition, each machine costs $c_i = c$, and our goal is to maximize the total profit (or equivalently, minimize the number of machines), while scheduling all the jobs and satisfying the capacity constraints. Note that we consider a single dimensional problem, where each job has one capacity requirement (e.g., the number of virtual CPU cores \emph{or} the amount of memory). Although cloud virtual machine packing may be modeled as a low-dimensional vector bin packing problem (see for example, \cite{microsoft2011validating}), one resource is often effectively binding and/or more critical so that focusing on it offers a much larger opportunity for overcommitment\footnotemark[4]\footnotetext[4]{Insofar as many vector bin packing heuristics are actually straightforward generalizations of the FF, NF and BF rules, it will become obvious how our proposed algorithm could similarly be adapted to the multi-resource setting in Section \ref{constantapprox}, although we do not pursue this idea in this paper.}.

\subsection{Bin packing problem}
For the case where $A_j$ is  deterministic, we obtain the classical \emph{deterministic bin packing} problem:
\begin{equation}
\label{equation:binpack:deter}
\tag{DBP}
  \begin{array}{ll}
        \displaystyle B = \min_{x_{ij}, y_i}
      & \displaystyle \sum_{i=1}^N y_i
    \\[12pt]
    \text{s.t.}
      & \displaystyle \sum_{j=1}^N A_j x_{ij} \leq V y_i 
          \qquad \forall i
    \\[11pt]
      & \displaystyle \sum_{i=1}^N x_{ij} = 1 \qquad \forall j
    \\[11pt]
      & x_{ij} \in \{0,1\} \qquad \forall i, j
          \\[11pt]
      & y_i \in \{0,1\} \qquad \forall i
  \end{array}
\end{equation}

For the offline version, we have a total of $N$ jobs and we need to decide which machines to use/purchase (captured by the decision variable $y_i$ that is equal to 1,
if machine $i$ is purchased and 0 otherwise). The solution is a $B$-partition of the set $\{1, 2, \ldots, N\}$ that satisfies the capacity constraints. The decision variable $x_{ij}$ equals one  if job $j$ is assigned to machine $i$ and zero otherwise.
As we discussed in Section \ref{lit}, there is an extensive literature on the DBP problem and its many variations covering both exact algorithms as well and approximation heuristics with performance bounds. 

The problem faced by a cloud provider is typically \emph{online} in nature since jobs arrive and depart over time. Unfortunately, 
it is not possible to continually re-solve the DBP problem as the data is updated for both practical and computational reasons. 
Keeping with the majority of prior work, we start by basing our
algorithms on static, single-period optimization formulations like the DBP problem, 
rather than explicitly modeling arrivals and departures. 
The next section explains how, unlike prior work, 
our single-period optimization model efficiently captures
the uncertainty faced by a cloud provider. 
We will consider both the online and offline versions of our model.

We remark that, while our online analysis considers sequentially arriving jobs, 
none of our results explicitly considers \emph{departing} jobs.
This is also in line with the bin-packing literature, where results usually apply to very 
general item arrival processes $\{ A_j \}$, but it is typically 
assumed that packed items remain in their assigned bins.
In practice, a large cloud provider is likely to be interested in a steady-state
where the distribution of jobs in the systems is stable over time (or at least predictable),
even if individual jobs come and go. 
Whereas the online model with arrivals correctly reflects that the scheduler cannot
optimize to account for unseen future arrivals,
it is unclear if and how additionally modeling departures would affect a system where 
the overall distribution of jobs remains the same over time. 
We therefore leave this question open.
Note that several works consider bin-packing with item departures (see, e.g., \cite{stolyar2015asymptotic} and the references therein). In this work, the authors design a simple greedy algorithm for general packing constraints and show that it can be asymptotically optimal.

\subsection{Chance constraints}\label{chanceconst}

The DBP problem suffers from the unrealistic assumption that the jobs' sizes $A_j$ are deterministic. In reality, jobs' requirements (or durations) can be highly unpredictable and quite volatile, especially from the perspective of a cloud provider with no control over the software executed in a virtual machine. Ensuring that the capacity constraints are satisfied for any realization of $A_j$ generally yields a conservative outcome. For example, if the jobs' true requirements are Bernoulli random variables taking on either 0.3 or 1.0 with equal probability, one needs to plan as if each job consumes a capacity of 1.0. By overcommitting resources, the provider can reduce the cost significantly. Caution is required however, since overcommitting can be very expensive if not done properly. Planning according to the expected value (in the previous simple example, 0.65), for instance, would result in capacity being too tight on many machines.
Specifically, for large machines, the realized requirements could exceed capacity up to half of the time. Depending on the specific resource and the degree of violation, such performance could be catastrophic for a cloud service provider. Concretely, sustained CPU contention among virtual machines would materially affect customers' performance metrics, whereas a shortage of available memory could require temporarily ``swapping'' some data to a slower storage medium with usually devastating consequences on performance. Other mitigations are possible, including migrating a running virtual machine to another host, but these also incur computational overhead for the provider and performance degradation for the customer. In the extreme case where overly optimistic scheduling results in inadequate capacity planning, there is even a stock-out risk where it is no longer possible to schedule all customers' jobs within a data center.
With this motivation in mind, our goal is to propose a formulation that finds the right overcommitment policy. We will show that by slightly overcommitting (defined formally in Section \ref{overcom}), one can reduce the costs significantly while satisfying the capacity constraints with high probability. 

While not strictly required by our approach, in practice, there is often an upper bound on $A_j$, denoted by $\bar {A_j}$. In the context of cloud computing, $\bar {A_j}$ is the requested capacity that a virtual machine is not allowed to exceed (32 CPU cores, or 128 GB of memory, say). However, the job may end up using much less, at least for some time. If the cloud provider schedules all the jobs according to their respective upper bounds $\bar {A_j}$, then there is no overcommitment. If the cloud provider schedules all the jobs according to some sizes smaller than the $\bar {A_j}$, then some of the machines may be overcommitted.

We propose to solve a bin packing problem with capacity chance constraints. Chance constraints are widely used in optimization problems, starting with \cite{charnes1963deterministic} for linear programs, and more recently in convex optimization (see, e.g., \cite{nemirovski2006convex}) and in finance (see, e.g., \cite{abdelaziz2007multi}). In this case,  the capacity constraints are replaced by:
\begin{align}
\label{equation:binpack:machineconstraint}
\mathbb{P} \Big (\sum_{j=1}^N A_j x_{ij} \leq V y_i \Big ) \geq \alpha,
\end{align}
where $\alpha$ represents the confidence level of satisfying the
constraint ($\alpha = 0.999$, say) and is exogenously set by the
cloud provider depending on considerations such as typical job's running time
and contractual agreements. Note that when $\alpha=1$, this
corresponds to the setting with no overcommitment, or in other
words, to the worst-case solution that covers all possible realizations of all the $A_j$'s. One of our goals is to study the trade-off between 
the probability of violating physical capacity and the cost reduction resulting from a given value of $\alpha$.

The problem becomes the \emph{bin packing with chance constraints}, parameterized by $\alpha$:
\begin{equation} 
\label{equation:binpack:single}
\tag{BPCC}
  \begin{array}{ll}
        \displaystyle B(\alpha) = \min_{x_{ij}, y_i}
      & \displaystyle \sum_{i=1}^N y_i
    \\[12pt]
    \text{s.t.}
      & \displaystyle \mathbb{P} \Big (\sum_{j=1}^N A_j x_{ij} 
             \leq V y_i \Big ) \geq \alpha \qquad \forall i
    \\[11pt]
      & \displaystyle \sum_{i=1}^N x_{ij} = 1 \qquad \forall j
    \\[11pt]
      & x_{ij} \in \{0,1\} \qquad \forall i, j
          \\[11pt]
      & y_i \in \{0,1\} \qquad \forall i
  \end{array}
\end{equation}

\subsection{Overcommitment}\label{overcom}
One can define the overcommitment level as follows. Consider two possible (equivalent) benchmarks. First,
one can solve the problem for $\alpha = 1$, and obtain a solution (by directly solving the IP or any other heuristic method) with objective $B(1)$. Then, we solve the problem for the desired value $\alpha < 1$. The overcommitment benefit can be defined as $0 < B(\alpha) / B(1) \leq 1$. It is
also interesting to compare the two different jobs assignments.

The second definition goes as follows. We define the \emph{overcommitment factor} as the amount of
\emph{sellable capacity} divided by the \emph{physical capacity of machines} in the data center, that is:
$$
 OCF(\alpha)
  \triangleq  \frac{\sum_j \bar{A}_j}{\sum_i V y_i}.
$$
Since we assume that all the machines have the same capacity and cost,
we can write:
\[OCF(\alpha)
  =   \frac{\sum_j \bar{A}_j}{VB(\alpha) }
  \ge \frac{\sum_j \bar{A}_j}{VB(1)}
  = OCF(1).
\]
Note that OCF(1) is (generally strictly) less than one, as the \emph{bin packing overhead} prevents the sale of all resources\footnotemark[5]\footnotetext[5]{Technical note: other production overheads such as safety stocks for various types of outages and management overheads, are generally also included in the denominator. For the purpose of this paper, we omit them.}.
Then, we have: 
\[
\frac{OCF(1)}{OCF(\alpha)} = \frac{B(\alpha)}{B(1)}.
\]
For illustration purposes, consider the following simple example with $N=70$ 1-core jobs. The jobs are independent and Bernoulli distributed with probability 0.5. In particular, the jobs are either high usage (i.e., fully utilize the 1 core), or low usage (in this case, idle). Each machine has a capacity $V=48$ cores. Without overcommitting, we need 2 machines, i.e., $B(1) = 2$. What happens if we schedule all the jobs in a single machine? In this case, one can reduce the cost (number of machines) by half, while satisfying the capacity constraint with probability 0.9987. In other words, $B(0.99) = 1$.
The overcommitment benefit in this simple example is clear. Our goal is to formalize a systematic way to overcommit in more complicated and realistic settings.

Note that overcommitment may lead to Service Level Agreement (SLA) violations. This paper does not discuss in detail the SLAs (with some possible associated metrics), and the corresponding estimation/forecast procedures as they are usually application and resource specific. Instead, this research treats a general Virtual Machine (VM) scheduling problem. More precisely, our context is that of a cloud computing provider with limited visibility into the mix of customer workloads, and hard SLAs. While the provider does track numerous service-level indicators, they are typically monotonic in the resource usage on average (we expect more work to translate to worse performance). Therefore, we believe that it is reasonable to rely on resource utilization as the sole metric in the optimization problem.

 \subsection{A variant of submodular bin packing}\label{altform}
 
In this section, we propose an alternative formulation that is closely related to the (BPCC) problem. Under some mild assumptions, we show that the latter is either exactly or approximately equivalent to the following \emph{submodular bin packing} problem:
\begin{equation}
\label{equation:binpack:single}
\tag{SMBP}
  \begin{array}{ll}
        \displaystyle B_S(\alpha) = \min_{x_{ij}, y_i}
      & \displaystyle \sum_{i=1}^N y_i
    \\[12pt]
    \text{s.t.}
      &  \sum_{j=1}^N \mu_j x_{ij} + D(\alpha) \sqrt{\sum_{j=1}^N b_j x_{ij}}
             \leq V y_i   \qquad \forall i
    \\[11pt]
      & \displaystyle \sum_{i=1}^N x_{ij} = 1 \qquad \forall j
    \\[11pt]
      & x_{ij} \in \{0,1\} \qquad \forall i, j
          \\[11pt]
      & y_i \in \{0,1\} \qquad \forall i
  \end{array}
\end{equation}
 
The difference between the (BPCC) and the (SMBP) problems is the way the capacity constraints are written. Here, we have replaced each chance constraint with a linear term plus a square root term. These constraints are submodular with respect to the vector $\mathbf x$. The variable $\mu_j$ denotes the expected value of $A_j$. In what follows, we will consider different definitions of $b_j$ and $D(\alpha)$ in three different settings. The first two are concrete motivational examples, whereas the third one is a generalization. In each case, we formally show the relation between the (BPCC) and the (SMBP) problems. 
 
\begin{enumerate}
\item \textbf{Gaussian case:} Assume that the random variables $A_j$ are Gaussian and independent. In this case, the random variable $Z = \sum_{j=1}^N A_j
x_{ij}$ for any given binary vector $\mathbf x$ is Gaussian,  and
therefore, one can use the following simplification:
\begin{align*}
\mathbb{P} \Big (\sum_{j=1}^N A_j x_{ij} \leq V y_i \Big ) = \mathbb{P} \Big (Z \leq V y_i \Big ) \geq \alpha.
\end{align*}
For each machine $i$, constraint \eqref{equation:binpack:machineconstraint} becomes:
\begin{align}\label{chance:gauss}
\sum_{j=1}^N \mu_j x_{ij} +  \Phi^{-1} (\alpha) \cdot  \sqrt{\sum_{j=1}^N \sigma^2_j x_{ij}} \leq V y_i,
\end{align}
where $\Phi^{-1}(\cdot)$ is the inverse CDF of a normal $N(0,1)$,
$\mu_j = \mathbb{E} [A_j]$ and $\sigma^2_j = \text{Var} (A_j)$. Note that we
have used the fact that $\mathbf x$ is binary so that $x^2_{ij} =
x_{ij}$. Consequently, the (BPCC) and the (SMBP) problems are equivalent with the values $b_j = \sigma^2_j$ and $D(\alpha) = \Phi^{-1}(\alpha)$.

When the random variables $A_j$ are independent but not normally distributed, if there are a large number of jobs per machine, one can apply the Central Limit Theorem and obtain a similar approximate argument. In fact, using a result from \cite{calafiore2006distributionally}, one can extend this equivalence to any radial distribution\footnotemark[6]\footnotetext[6]{Radial distributions include all probability densities whose level sets are ellipsoids.  The formal mathematical definition can be found in \cite{calafiore2006distributionally}.}.
 
 \item \textbf{Hoeffding's inequality:} Assume that the random variables $A_j$ are independent with a finite support $[ \underline{A}_j , \overline{A}_j  ]$, $0 \leq \underline{A}_j < \overline{A}_j $ with mean $\mu_j$. As we discussed, one can often know the value of $\overline{A}_j$ and use historical data to estimate $\mu_j$ and $\underline{A}_j$ (we discuss this in more detail in Section \ref{comps}). Assume that the mean usages fit on each machine, i.e.,  $\sum_{j=1}^N x_{ij} \mu_j < y_i V_i$. Then, Hoeffding's inequality states that:
\begin{align*}
 \displaystyle \mathbb{P} \Big (\sum_{j=1}^N A_j x_{ij} 
             \leq V y_i \Big ) \geq 1 - e^{\frac{-2  [ V y_i - \sum_{j=1}^N \mu_j x_{ij}  ]^2} {\sum_{j=1}^N (\overline{A}_j - \underline{A}_j)^2 } }.
\end{align*}
Equating the right hand side to $\alpha$, we obtain:
 \begin{align*}
\frac{-2  [ V y_i - \sum_{j=1}^N \mu_j x_{ij} ]^2} {\sum_{j=1}^N b_j x_{ij} } = \ln (1 - \alpha),
\end{align*}
where $b_j = (\overline{A}_j - \underline{A}_j)^2$ represents the range of job $j$'s usage.  Re-arranging the equation, we obtain for each machine $i$:
\begin{align}\label{HIapprox}
\sum_{j=1}^N \mu_j x_{ij}+ D(\alpha) \sqrt{ \sum_{j=1}^N b_j x_{ij}} \leq V y_i,
\end{align}
where in this case, $D(\alpha) = \sqrt{-0.5 \ln(1 - \alpha)}$. Note that in this setting the (BPCC) and the (SMBP) problems are not equivalent.
We only have that any solution of the latter is a feasible solution for the former. We will demonstrate in Section \ref{comps} that despite being very conservative, this formulation based on Hoeffding's inequality actually yields good practical solutions.

The next case is a generalization of the last two.

\item \textbf{Distributionally robust formulations:} Assume that the random variables $A_j$ are independent with some unknown distribution. We only know that this distribution belongs to a family of probability distributions $\mathcal D$. We consider two commonly used examples of such families. First, we consider the family $\mathcal D_1$ of distributions with a given mean and (diagonal) covariance matrix, $\mathbf{\mu}$ and $\mathbf{\Sigma}$, respectively. Second, we look at $\mathcal D_2$, the family of generic distributions of independent random variables over bounded intervals $[ \underline{A}_j , \overline{A}_j  ]$. 


In this setting, the chance constraint is assumed to be enforced robustly with respect to the entire family $\mathcal D$ of probability distributions on $\mathbf A = (A_1, A_2, \ldots, A_N)$, meaning that:
 \begin{align}\label{robustchance}
\inf_{\mathbf A \sim \mathcal{D}} \mathbb{P} \Big (\sum_{j=1}^N A_j x_{ij} \leq V y_i \Big ) \geq \alpha.
\end{align}

In this context, we have the following result.

\begin{proposition}\label{robustSMBP}
Consider the robust bin packing problem with the capacity chance constraints \eqref{robustchance} for each machine $i$. Then, for any $\alpha \in (0,1)$, we have:
\begin{itemize}

\item For the family $\mathcal D_1$ of distributions with a given mean and diagonal covariance matrix, the robust problem is equivalent to the (SMBP) with $b_j = \sigma^2_j$ and $D_1(\alpha) = \sqrt{\alpha/(1-\alpha)}$. 

\item For the family $\mathcal D_2$ of generic distributions of independent random variables over bounded intervals, the robust problem can be approximated by the (SMBP) with $b_j = (\overline{A}_j - \underline{A}_j)^2$ and $D_2(\alpha) = \sqrt{-0.5 \ln(1 - \alpha)}$.

\end{itemize}
\end{proposition}
\end{enumerate}
The details of the proof are omitted for conciseness. In particular, the proof for $\mathcal D_1$ is analogous to an existing result in continuous optimization that converts linear programs with a chance constraint into a linear program with a convex second-order cone constraint (see \cite{calafiore2006distributionally} and \cite{ghaoui2003worst}). The proof for $\mathcal D_2$ follows directly from the fact that Hoeffding's inequality applies for all such distributions, and thus for the infimum of the probability.

We have shown that the (SMBP) problem is a good approximation for the bin packing problem with chance constraints. For the case of independent random variables with a given mean and covariance, the approximation is exact and for the case of distributions over independent bounded intervals, the approximation yields a feasible solution. We investigate practical settings in Section \ref{comps}, and show that these approximate formulations all yield good solutions to the original problem. From now on, we consider solving the (SMBP) problem, that is repeated here for convenience:
 \begin{equation}
\label{equation:binpack:single}
\tag{SMBP}
  \begin{array}{ll}
        \displaystyle B_S(\alpha) = \min_{x_{ij}, y_i}
      & \displaystyle \sum_{i=1}^N y_i
    \\[12pt]
    \text{s.t.}
      &  \sum_{j=1}^N \mu_j x_{ij} + D(\alpha) \sqrt{\sum_{j=1}^N b_j x_{ij}}
             \leq V y_i   \qquad \forall i
    \\[11pt]
      & \displaystyle \sum_{i=1}^N x_{ij} = 1 \qquad \forall j
    \\[11pt]
      & x_{ij} \in \{0,1\} \qquad \forall i, j
          \\[11pt]
      & y_i \in \{0,1\} \qquad \forall i
  \end{array}
\end{equation}
As discussed, the capacity constraint is now replaced by the following equation, called the \emph{modified capacity constraint}:
\begin{align}\label{general:sub}
\sum_{j=1}^N \mu_j x_{ij}+ D(\alpha) \sqrt{ \sum_{j=1}^N b_j x_{ij}} \leq V y_i.
\end{align}
One can interpret equation \eqref{general:sub} as follows. Each machine has a capacity $V$. Each job $j$ consumes capacity $\mu_j$ in expectation, as well as an additional buffer to account for the uncertainty. This buffer depends on two factors: (i) the variability of the job, captured by the parameter $b_j$; and (ii) the acceptable level of risk through $D(\alpha)$. The function $D(\alpha)$ is increasing in $\alpha$, and therefore we impose a stricter constraint as $\alpha$ approaches 1  by requiring this extra buffer to be larger.

Equation \eqref{general:sub} can also be interpreted as a risk measure applied by the scheduler. For each machine $i$, the total (random) load is $\sum_{j=1}^N A_j x_{ij}$. If we consider that $\mu_j$ represents the expectation and $b_j$ corresponds to the variance, then $\sum_{j=1}^N \mu_j x_{ij}$ and $\sqrt{ \sum_{j=1}^N b_j x_{ij}}$ correspond to the expectation and the standard deviation of the total load on machine $i$ respectively. As a result, the right hand side of equation \eqref{general:sub} can be interpreted as an adjusted risk utility, where $D(\alpha)$ is the degree of risk aversion of the scheduler. The additional amount allocated for job $j$ can be interpreted as a safety buffer to account for the uncertainty and for the risk that the provider is willing to bear. As we discussed, this extra buffer decreases with the number of jobs assigned to the same machine. 
%
In Section \ref{constantapprox}, we develop efficient methods to solve the (SMBP) with analytical performance guarantees.

\subsection{Two naive approaches}\label{naive}

In this section, we explore the limitations of two approaches that come to mind. The first attempt is to rewrite the problem as a linear integer program: the decision variables are all binary and the non-linearity in (SMBP) can actually be captured by common modeling techniques, as detailed in Appendix \ref{append:naive}. Unfortunately, solving this IP is not a viable option. Similarly as for the classical deterministic bin packing problem, solving even moderately large instances with commercial solvers takes several hours. Moreover, applying the approach to smaller, specific toy instances provides little insight about the assignment policy, and how the value of $\alpha$ affects the solution. Since our goal is to develop practical strategies for the online problem, we chose not to further pursue exact solutions.

The second potential approach is to develop an algorithm for a more general problem: the bin packing with general monotone submodular capacity constraints. Unfortunately, using some machinery and results from \cite{goemans2009approximating} and \cite{SvitkinaFleischer}, we next show that it is in fact impossible to find a solution within any reasonable factor from optimal. 

\begin{theorem}\label{general:submod}
Consider the bin packing problem with general monotone submodular capacity constraints for each machine. Then, it is impossible to guarantee a solution within a factor better than $\frac{\sqrt{N}}{\ln(N)}$ from optimal.
\end{theorem}
The proof can be found in Appendix \ref{append:naive}. We will show that the (SMBP) problem that we consider is more tractable as it concerns only a specific class of monotone submodular capacity constraints that capture the structure of the chance-constrained problem.
In the next session, we start by addressing simple special cases in order to draw some structural insights.
\section{Results and insights for special cases}\label{analysis}

In this section, we consider the (SMBP) problem for some given $\mu_j$, $b_j$, $N$ and $D(\alpha)$. Our goals are to: (i) develop efficient approaches to solve the problem; (ii) draw some insights on how to schedule the different jobs and; (iii) study the effect of the different parameters on the outcome. This will ultimately allows us to understand the impact of overcommitment in resource allocation problems, such as cloud computing. 

\subsection{Identical distributed jobs}

\begin{figure}[h!]
\begin{center}
\includegraphics[width=3.1in]{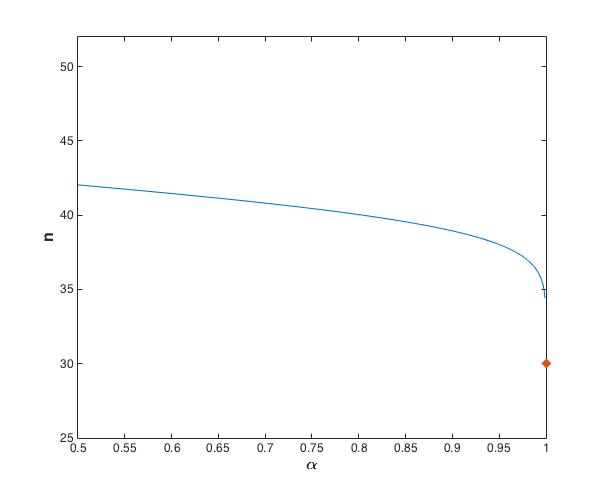}
\end{center}
\caption{Parameters: $\overline{A}=1$, $\underline{A}=0.3$, $\mu=0.65$, $V=30$}\label{N:alpha:symmetric}
\end{figure}

We consider the symmetric setting where all the random variables $A_j$ have the same distribution, such that $\mu_j  = \mu $ and $b_j = b$ in the (SMBP) problem. By symmetry, we only need to find the number of jobs $n$ to assign to each machine. Since all the jobs are interchangeable, our goal is to assign as many jobs as possible in each machine. In other words, we want to pick the largest value of $n$ such that the constraint \eqref{general:sub} is satisfied, or equivalently:
\begin{align*}  
n \mu + D(\alpha) \sqrt{n b} &\le V  \\
D(\alpha)^2 &\le \frac{\lbrack V - n \mu \rbrack^2}{n b }.
\end{align*}
For a given value of $\alpha$, this is the largest integer smaller than:
\begin{align}\label{N:alpha}
n (\alpha) = \frac{V} {\mu} + \frac{1} {2 \mu^2} \big [ b D(\alpha)^2  - \sqrt{b^2 D(\alpha)^4 + 4 b D(\alpha)^2 V \mu} \big].
\end{align}
\begin{itemize}
\item For a given value of $\alpha$, the number of jobs $n (\alpha)$ increases with $V/\mu$. Indeed, since $\mu$ represents the expected job size, increasing the ratio $V/\mu$ is equivalent to increasing the number of "average" jobs a machine can host. If the jobs are smaller or the machines larger, one can fit more jobs per machine, as expected.

\item For a given value of $V/\mu$,  $n (\alpha)$ is a non-increasing function of $\alpha$. When $\alpha$ increases, it means that we enforce the capacity constraint in a stricter manner (recall that $\alpha = 1$ corresponds to the case without overcommitment). As a result, the number of jobs per machine cannot increase. 

\item For given values of $\alpha$ and $V/\mu$,  $n (\alpha)$ is a non-increasing function of $b$. Recall that the parameter $b$ corresponds to some measure of spread (the variance in the Gaussian setting, and the range for distributions with bounded support). Therefore, when $b$ increases, it implies that the jobs' resource usage is more volatile and hence, a larger buffer is needed. Consequently, the number of jobs cannot increase when $b$ grows. 

\item For given values of $\alpha$ and $V$, $n (\alpha)$ is non-increasing with $\sqrt b/\mu$. The quantity $\sqrt b/\mu$ represents the coefficient of variation of the random job size in the Gaussian case, or a similarly normalized measure of dispersion in other cases. Consequently, one should be able to fit less jobs, as the variability increases.
 
\end{itemize}
The simple case of identically distributed jobs allows us to understand how the different factors affect the number of jobs that one can assign to each machine. In Figure \ref{N:alpha:symmetric}, we plot equation \eqref{N:alpha} for an instance with $\overline{A}=1$, $\underline{A}=0.3$, $\mu=0.65$, $V=30$ and $0.5 \leq \alpha < 1$. The large dot for $\alpha=1$ in the figure represents the case without overcommitment (i.e., $\alpha = 1$). Interestingly, one can see that when the value of $\alpha$ approaches 1, the benefit of allowing a small probability of violating the capacity constraint is significant, so that one can increase the number of jobs per machine. In this case, when $\alpha = 1$, we can fit 30 jobs per machine, whereas when $\alpha = 0.992$, we can fit 36 jobs, hence, an improvement of 20\%.
Note that this analysis guarantees that the capacity constraint is satisfied with at least probability $\alpha$. As we will show in Section \ref{comps} for many instances, the capacity constraint is satisfied with an even higher probability. 

\begin{figure}[h!]
\centering
  \begin{subfigure}[b]{0.43\textwidth}
\includegraphics[width=3.1in]{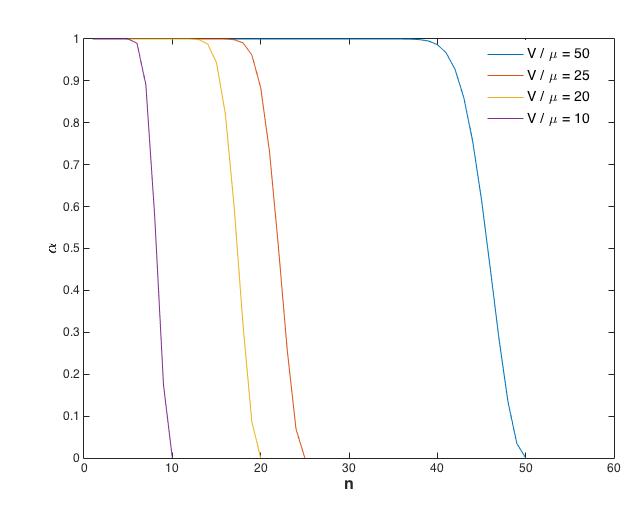}
\caption{Parameters: $\overline{A}=1$, $\underline{A}=0.3$, $\mu=0.65$}\label{alpha:N:symmetric}
  \end{subfigure}
  \hspace{0.02\textwidth}
  \begin{subfigure}[b]{0.43\textwidth}
  \includegraphics[width=3.2in]{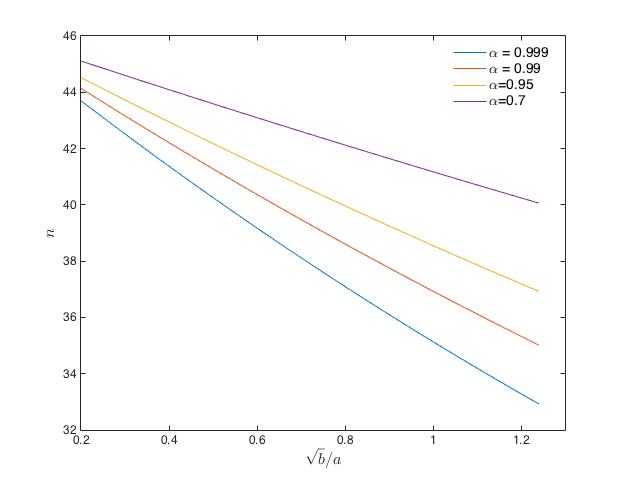}
\caption{Parameters: $V = 30$}\label{coeff:symmetric}
  \end{subfigure}
\caption{Example for identically distributed jobs}
\label{fig:homogeneous}
\end{figure}

Alternatively, one can plot $\alpha$ as a function of $n$ (see Figure \ref{alpha:N:symmetric} for an example with different values for $V / \mu$). As expected, the benefit of overcommitting increases with $V / \mu$, i.e., one can fit a larger number of jobs per machine. In our example, when $V / \mu =25$, by scheduling jobs according to $\overline{A}$ (i.e., $\alpha=1$, no overcommitment), we can schedule 14 jobs, whereas if we allow a 0.1\% violation probability, we can schedule 17 jobs. Consequently, by allowing 0.1\% chance of violating the capacity constraint, one can save more than 20\% in costs. 

We next discuss how to solve the problem for the case with a small number of different classes of job distributions.

\subsection{Small number of job distributions}

We now consider the case where the random variables $A_j$ can be clustered in few different categories. For example, suppose standard clustering algorithms are applied to historical data to treat similar jobs as a single class with some distribution of usage. For example, one can have a setting with four types of jobs: (i) large jobs with no variability ($\mu_j$ is large and $b_j$ is zero); (ii) small jobs with no variability ($\mu_j$ is small and $b_j$ is zero);  (ii) large jobs with high variability (both $\mu_j$ and $b_j$ are large); and (iv) small jobs with high variability ($\mu_j$ is small and $b_j$ is high). In other words, we have $N$ jobs and they all are from one of the 4 types, with given values of $\mu_j$ and $b_j$. The result for this setting is summarized in the following Observation (the details can be found in Appendix \ref{CSP:details}). 
\begin{obs}\label{CSP:obs}
In the case where the number of different job classes is not too large, one can solve the problem efficiently as a cutting stock problem. 
\end{obs}
The resulting cutting stock problem (see formulation \eqref{equation:cuttingstock:1D} in Appendix \ref{CSP:details}) is well studied in many contexts (see \cite{gilmore1961linear} for a classical approach based on linear programming, or the recent survey of \cite{delorme2016exactcsp}). For example, one can solve the LP relaxation of \eqref{equation:cuttingstock:1D} and round the fractional solution. This approach can be very useful for cases where the cloud provider have enough historical data, and when the jobs can all be regrouped into a small number of different clusters. This situation is sometimes realistic but not always. Very often, grouping all possible customer job profiles into a small number of classes, each described by a single distribution is likely unrealistic in many contexts. For example, virtual machines are typically sold with 1, 2, 4, 8, 16, 32 or 64 CPU cores, each with various memory configurations, to a variety of customers with disparate use-cases. Aggregating across these jobs is already dubious, before considering differences in their usage means and variability. Unfortunately, if one decides to use a large number of job classes, solving a cutting stock problem is not scalable. In addition, this approach requires advance knowledge of the number of jobs of each class and hence, cannot be applied to the online version of our problem.


\section{Online constant competitive algorithms}\label{constantapprox}
In this section, we analyze the performance of a large class of algorithms for the online version of problem (SMBP). We note that the same guarantees hold for the offline case, as it is just a simpler version of the problem. We then present a refined result for the offline problem in Section \ref{offline}. 

\subsection{Lazy algorithms are $\frac{8}{3}$-competitive}

An algorithm is called \emph{lazy}, if it does not purchase/use a new machine unless necessary. The formal definition is as follows.

\begin{definition}
We call an online algorithm \emph{lazy} if upon arrival of a new job, it assigns the job to one of the existing (already purchased) machines given the capacity constraints are not violated. In other words, the algorithm purchases a new machine if and only if non of the existing machines can accomodate the newly arrived job. 
\end{definition} 

Several commonly used algorithms fall into this category, e.g., First-Fit, Best-Fit, Next-Fit, greedy type etc. 
Let $\opt$ be the optimal objective, i.e., the minimum number of machines needed to serve all the jobs $\{1, 2, \cdots, N\}$. Recall that in our problem, each job $1 \leq j \leq N$, has two characteristics: $\mu_j$ and $b_j$ which represent the mean and the uncertain part of job $j$ respectively.  
For a set of jobs $S$, we define the corresponding cost $\cost(S)$ to be $\sum_{j \in S} \mu_j + \sqrt{\sum_{j \in S} b_j}$. 
Without loss of generality, we can assume (by normalization of all $\mu_j$ and $b_j$) that the capacity of each machine is $1$ and that $D(\alpha)$ is also normalized to 1.
We call a set $S$ feasible, if its cost is at most the capacity limit $1$.
In the following Theorem, we show that any lazy algorithm yields a constant approximation for the (SBBP) problem. 

\begin{theorem}\label{thm:lazyapp}
Any lazy algorithm $\alg$ purchases at most $\frac{8}{3}\opt$ machines, where $\opt$ is the optimum number of machines to serve all jobs. 
\end{theorem}

\proof{Proof.}
Let $m$ be the number of machines that $\alg$ purchases when serving jobs $\{1, 2, \cdots, N\}$. For any machine $1 \leq i \leq m$, we define $S_i$ to be the set of jobs assigned to machine $i$. Without loss of generality, we assume that the $m$ machines are purchased in the order of their indices. In other words, machines $1$ and $m$ are the first and last purchased ones respectively. 

For any pair of machines $1 \leq i <  i' \leq m$, we next prove that the set $S_i \cup S_{i'}$ is infeasible for the modified capacity constraint, i.e., $\sum_{j \in S_i \cup S_{i'}} \mu_j + \sqrt{\sum_{j \in S_i \cup S_{i'}} b_j} > 1$. Let $j$ be the first job assigned to machine $i'$. Since $\alg$ is lazy, assigning $j$ to machine $i$ upon its arrival time was not feasible, i.e., the set $\{j\} \cup S_i$ is infeasible. Since we only assign more jobs to machines throughout the course of the algorithm, and do not remove any job, the set $S_{i'} \cup S_i$ is also infeasible. 

In the next Lemma, we lower bound the sum of $\mu_j+b_j$ for the jobs in an infeasible set. 
\begin{lemma}\label{lem:lowerbound:34}
For any infeasible set $T$, we have $\sum_{j \in T} (\mu_j + b_j) > \frac{3}{4}$.
\end{lemma}
\proof{Proof.}
For any infeasible set $T$, we have by definition $\sum_{j \in T} \mu_j + \sqrt{\sum_{j \in T} b_j} > 1$. We denote $x = \sum_{j \in T} \mu_j$, and $y = \sqrt{\sum_{j \in T} b_j} $. Then, $y > 1 -  x$. 
If $x$ is greater than $1$, the claim of the lemma holds trivially. Otherwise, we obtain:
$$
x + y^2 > x + (1-x)^2 = x^2 - x + 1 = (x - \frac{1}{2})^2 + \frac{3}{4} \geq \frac{3}{4}.
$$
We conclude that $\sum_{j \in T} (\mu_j + b_j) > \frac{3}{4}$. $\Box$ 
\endproof

As discussed, for any pair of machines $i < i'$,  the union of their sets of jobs $S_{i} \cup S_{i'}$ is an infeasible set that does not fit in one machine.
We now apply the lower bound from Lemma \ref{lem:lowerbound:34} for the infeasible set $S_{i} \cup S_{i'}$ and imply $\sum_{j \in S_{i} \cup S_{i'}} (\mu_j + b_j) > \frac{3}{4}$. 
We can sum up this inequality for all ${m \choose 2}$ pairs of machines $i$ and $i'$ to obtain: 
\begin{eqnarray}\label{eq:SumInfeasiblePairs}
\sum_{1 \leq i < i' \leq m} \sum_{j \in S_{i} \cup S_{i'}} (\mu_j + b_j) > \frac{3}{4}{m \choose 2}.
\end{eqnarray}

We claim that the left hand side of inequality~\eqref{eq:SumInfeasiblePairs} is equal to $(m-1) \sum_{j=1}^N (\mu_j + b_j)$. We note that for each job $j \in S_k$, the term $\mu_j + b_j$ appears $k-1$ times in the left hand side of inequality~\eqref{eq:SumInfeasiblePairs} when $i'$ is equal to $k$. In addition, this term also appears $m-k$ times when $i$ is equal to $k$. Therefore, every $\mu_j + b_j$ appears $k-1+m-k = m-1$ times, which is independent of the index $k$ of the machine that contains job $j$. 
As a result, we obtain:
$$
\sum_{j=1}^N (\mu_j + b_j) >
\frac{3}{4 (m-1)}{m \choose 2} = \frac{3m} {8}.
$$
On the other hand, we use the optimal assignment to upper bound the sum $\sum_{j=1}^N (\mu_j + b_j)$, and relate $m$ to $\opt$. Let $T_1, T_2, \cdots, T_{\opt}$ be the optimal assignment of all jobs to $\opt$ machines. Since $T_i$ is a feasible set, we have $\sum_{j \in T_i} \mu_j + \sqrt{\sum_{j \in T_i}b_j} \leq 1$, and consequently, we also have $\sum_{j \in T_i} (\mu_j + b_j) \leq 1$. Summing up for all the machines $1 \leq i \leq \opt$, we obtain: $\sum_{j=1}^N (\mu_j + b_j) = \sum_{i=1}^{\opt} \sum_{j \in T_i} (\mu_j + b_j) \leq \opt$. We conclude that:

$$
\opt \geq \sum_{j=1}^N (\mu_j + b_j) > \frac{3m}{8}.
$$
This completes the proof of $m < \frac{8\opt}{3}$. $\Box$ 
\endproof

Theorem \ref{thm:lazyapp} derives an approximation guarantee of $8/3$ for any lazy algorithm. In many practical settings, one can further exploit the structure of the set of jobs, and design algorithms that achieve better approximation factors. For example, if some jobs are usually larger relative to others, one can incorporate this knowledge into the algorithm. 
We next describe the main intuitions behind the $8/3$ upper bound. In the proof of Theorem~\ref{thm:lazyapp}, we have used the following two main proof techniques: 
\begin{itemize}
\item
First, we show a direct connection between the feasibility of a set $S$ and the sum $\sum_{j \in S} (\mu_j + b_j)$. In particular, we prove that $\sum_{j \in S} (\mu_j + b_j) \leq 1$ for any feasible set, and greater than $3/4$ for any infeasible set. Consequently, $\opt$ cannot be less than the sum of $\mu_j + b_j$ for all jobs. 
The gap of $4/3$ between the two bounds contributes partially to the final upper bound of $8/3$.
\item
Second, we show that the union of jobs assigned to any pair of machines by the lazy algorithm is an infeasible set, so that their sum of $\mu_j + b_j$ should exceed $3/4$. One can then find $m/2$ disjoint pairs of machines, and obtain a lower bound of $3/4$ for the sum $\mu_j + b_j$ for each pair. The fact that we achieve this lower bound for every pair of machines (and not for each machine) contributes another factor of $2$ to the approximation factor, resulting to $ \frac{4}{3} \times 2 = \frac{8}{3}$. 
\end{itemize}

Note that the second loss of a factor of $2$ follows from the fact that the union of any two machines forms an infeasible set and nothing stronger. In particular, all machines could potentially have a cost of $1/2+\epsilon$ for a very small $\epsilon$, and make the above analysis tight. Nevertheless, if we assume that each machine is nearly full (i.e., has $\cost$ close to $1$), one can refine the approximation factor.

\begin{theorem}\label{thm:lazybetterapp}
For any $0 \leq \epsilon \leq 0.3$, if the lazy algorithm $\alg$  assigns all the jobs to $m$ machines such that $\cost(S_i) \geq 1-\epsilon$ for every $1 \leq i \leq m$, we have $m \leq (\frac{4}{3}+3\epsilon)\opt$, i.e., a $(\frac{4}{3}+3\epsilon)$ approximation guarantee. 
\end{theorem}

\proof{Proof.}
To simplify the analysis, we denote $\beta$ to be $1-\epsilon$.
For a set $S_i$, we define $x = \sum_{j \in S_i} \mu_j$ and $y = \sqrt{\sum_{j \in S_i} b_j}$.
Since $\cost(S_i)$ is at least $\beta$, we have $x + y \geq \beta$. Assuming $x \leq \beta$, we have:
$$
\sum_{j \in S_i} (\mu_j + b_j) = x+y^2 \geq x + (\beta -x)^2 = \Big ( x - \frac{2\beta-1}{2} \Big)^2 + \beta - \frac{1}{4} \geq \beta - \frac{1}{4} = \frac{3}{4} - \epsilon,
$$ 
where the first equality is by the definition of $x$ and $y$, the second inequality holds by $x+y \geq \alpha$, and the rest are algebraic manipulations. 
For $x > \beta$, we also have $\sum_{j \in S_i} (\mu_j + b_j) \geq x > \beta > \frac{3}{4} - \epsilon$.
We conclude that $\sum_{j=1}^N (\mu_j + b_j) \geq m \times (\frac{3}{4} - \epsilon)$. We also know that $\opt \geq \sum_{j=1}^N (\mu_j + b_j)$, which implies that $m \leq \frac{\opt}{3/4 - \epsilon} \leq \opt (\frac{4}{3}+3\epsilon)$ for $\epsilon \leq 0.3$.
$\Box$ 
\endproof

A particular setting where the condition of Theorem~\ref{thm:lazybetterapp} holds is when the capacity of each machine is large compared to all jobs, i.e., $\max_{1 \leq j \leq N} \mu_j+\sqrt{b_j}$ is at most $\epsilon$. In this case, for each machine $i \neq m$ (except the last purchased machine), we know that there exists a job $j \in S_m$ (assigned to the last purchased machine $m$) such that the algorithm could not assign $j$ to machine  $i$. This means that $\cost(S_i \cup \{j\})$ exceeds one. Since $\cost$ is a subadditive function, we have $\cost(S_i \cup \{j\}) \leq \cost(S_i) + \cost(\{j\})$. We also know that $\cost(\{j\}) \leq \epsilon$ which implies that $\cost(S_i) > 1-\epsilon$.

\begin{remark}
As elaborated above, there are two main sources for losses in the approximation factors: non-linearity of the cost function that can contribute up to $4/3$, and machines being only partially full that can cause an extra factor of $2$ which in total implies the $8/3$ approximation guarantee. 
In the classical bin packing case (i.e., $b_j = 0$ for all $j$), the cost function is linear, and the non-linearity losses in approximation factors fade. Consequently, we obtain that (i) Theorem \ref{thm:lazyapp} reduces to a 2 approximation factor; and (ii) Theorem \ref{thm:lazybetterapp} reduces to a $(1 + \epsilon)$ approximation factor, which are both consistent with known results from the literature on the classical bin packing problem.
\end{remark}

Theorem \ref{thm:lazybetterapp} improves the bound for the case where each machine is almost full. However, in practice machines are often not full. In the next section, we derive a bound as a function of the minimum number of jobs assigned to the machines.

\subsection{Algorithm $\firstfit$ is $\frac{9}{4}$-competitive}
So far, we considered the general class of lazy algorithms. One popular algorithm in this class (both in the literature and in practice) is $\firstfit$. By exploring the structural properties of allocations made by $\firstfit$, we can provide a better competitive ratio of $\frac{9}{4} < \frac{8}{3}$. Recall that upon the arrival of a new job, $\firstfit$ purchases a new machine if the job does not fit in any of the existing machines. Otherwise, it assigns the job to the first machine (based on a fixed ordering such as machine IDs) that it fits in. This algorithm is simple to implement, and very well studied in the context of the classical bin packing problem. First, we present an extension of Theorem \ref{thm:lazyapp} for the case where each machine has at least $K$ jobs.

\begin{corollary}\label{atleastKjobs}
If the $\firstfit$ algorithm assigns jobs such that each machine receives at least $K$ jobs, the number of purchased machines does not exceed $\frac{4}{3} (1 + \frac{1} {K}) \opt$, where $\opt$ is the optimum number of machines to serve all jobs. 
\end{corollary}
One can prove Corollary \ref{atleastKjobs} in a similar fashion as the proof of Theorem \ref{thm:lazyapp} and using the fact that jobs are assigned using $\firstfit$ (the details are omitted for conciseness). For example, when $K=2$ (resp. $K=5$), we obtain a 2 (resp. 1.6) approximation. We next refine the approximation factor for the problem by using the $\firstfit$ algorithm.

\begin{theorem}\label{firstfit94}
The number of purchased machines by Algorithm $\firstfit$ for any arrival order of jobs is not more than $\frac{9}{4}\opt + 1$. 
\end{theorem}

The proof can be found in Appendix \ref{proof94}. We note that the approximation guarantees we developed in this section do not depend on the factor $D(\alpha)$, and on the specific definition of the parameters $\mu_j$ and $b_j$. In addition, as we show computationally in Section \ref{comps}, the performance of this class of algorithm is not significantly affected by the factor $D(\alpha)$.

\section{Insights on job scheduling}\label{concav}

In this section, we show that guaranteeing the following two guidelines in any allocation algorithm yields optimal solutions:
\begin{itemize}
\item
Filling up each machine completely such that no other job fits in it, i.e., making each machine's $\cost$ equal to $1$.
\item
Each machine contains a set of \emph{similar} jobs (defined formally next).
\end{itemize}

We formalize these properties in more detail, and show how one can achieve optimality by satisfying these two conditions. We call a machine \emph{full} if $\sum_{j \in S} \mu_j + \sqrt{\sum_{j \in S} b_j}$ is equal to $1$ (recall that the machine capacity is normalized to 1 without loss of generality), where $S$ is the set of jobs assigned to the machine. Note that it is not possible to assign any additional job (no matter how small the job is) to a full machine. Similarly, we call a machine \emph{$\epsilon$-full}, if the the cost is at least $1 - \epsilon$, i.e., $\sum_{j \in S} \mu_j + \sqrt{\sum_{j \in S} b_j} \geq 1 - \epsilon$. 
We define two jobs to be \emph{similar}, if they have the same $b / \mu$ ratio. Note that the two jobs can have different values of $\mu$ and $b$.
We say that a machine is \emph{homogeneous}, if it only contains similar jobs. In other words, if the ratio $b_j / \mu_j$ is the same for all the jobs $j$ assigned to this machine. By convention, we define $b_j / \mu_j$ to be $+\infty$ when $\mu_j = 0$. In addition, we introduce the relaxed version of this property: we say that two jobs are \emph{$\delta$-similar}, if their $b / \mu$ ratios differ by at most a multiplicative factor of $1+\delta$. A machine is called \emph{$\delta$-homogeneous}, if it only contains $\delta$-similar jobs (i.e., for any pair of jobs $j$ and $j'$ in the same machine, $\frac{b_j/\mu_j}{b'_j/\mu'_j}$ is at most $1+\delta$).

\begin{theorem}\label{thm:full-homogeneous}
For any $\epsilon \geq 0$ and $\delta \geq 0$, consider an assignment of all jobs to some machines with two properties: a) each machine is $\epsilon$-full, and b) each machine is $\delta$-homogeneous. Then, the number of purchased machines in this allocation is at most $\frac{\opt}{(1-\epsilon)^2(1-\delta)}$.
\end{theorem}   
The proof can be found in Appendix \ref{proofopt}.

In this section, we proposed an easy to follow recipe in order to schedule jobs to machines. Each arriving job is characterized by two parameters $\mu_j$ and $b_j$. Upon arrival of a new job, the cloud provider can compute the ratio $r_j = b_j / \mu_j$. Then, one can decide of a few buckets for the different values of $r_j$, depending on historical data, and performance restrictions. Finally, the cloud provider will assign jobs with similar ratios to the same machines and tries to fill in machines as much as possible. In this paper, we show that such a simple strategy guarantees a good performance (close to optimal) in terms of minimizing the number of purchased machines while at the same time allowing to strategically overcommit.
\section{Extensions}\label{extens}

In this section, we present two extensions of the problem we considered in this paper. 

\subsection{Offline $2$-approximation algorithm}\label{offline}
Consider the offline version of the (SMBP) problem. In this case, all the $N$ jobs already arrived, and one has to find a feasible schedule so as to minimize the number of machines. We propose the algorithm $\localsearch$ that iteratively reduces the number of purchased machines, and also uses ideas inspired from $\firstfit$ in order to achieve a 2-approximation for the offline problem. 
Algorithm $\localsearch$ starts by assigning all the jobs to machines arbitrarily, and then iteratively refines this assignment. Suppose that each machine has a unique identifier number. 
We next introduce some notation before presenting the update operations. Let $a$ be the number of machines with only one job, $A_1$ be the set of these $a$ machines, and $S_1$ be the set of jobs assigned to these machines. Note that this set changes throughout the algorithm with the update operations. We say that a job $j \notin S_1$ is \emph{good}, if it fits in at least $6$ of the machines in the set $A_1$\footnotemark[7]\footnotetext[7]{The reason we need 6 jobs is technical, and will be used in the proof of Theorem \ref{offline2}.}. In addition, we say that a machine is \emph{large}, if it contains at least $5$ jobs, and we denote the set of large machines by $A_5$. We say that a machine is \emph{medium size}, if it contains $2$, $3$, or $4$ jobs, and we denote the set of medium machines by $A_{2,3,4}$. We call a medium size machine \emph{critical}, if it contains one job that fits in none of the machines in $A_1$, and the rest of the jobs in this machine are all good. Following are the update operations that $\localsearch$ performs until no such operation is available.

\begin{itemize}
\item
Find a job $j$ in machine $i$ ($i$ is the machine identifier number) and assign it to some other machine $i' < i$ if feasible (the outcome will be similar to $\firstfit$).
\item
Find a medium size machine $i$ that contains only good jobs. Let $j_1, \cdots, j_{\ell}$ ($2 \leq \ell \leq 4$) be the jobs in machine $i$. Assign $j_1$ to one of the  machines in $A_1$ that it fits in. Since $j_1$ is a good job, there are at least $6$ different options, and the algorithm picks one of them arbitrarily. Assign $j_2$ to a different machine in $A_1$ that it fits in. There should be at least $5$ ways to do so. We continue this process until all the jobs in machine $i$ (there are at most $4$ of them) are assigned to distinct machines in $A_1$, and they all fit in their new machines. This way, we release machine $i$ and reduce the number of machines by one. 
\item
Find a medium size machine $i$ that contains one job $j$ that fits in at least one machine in $A_1$, and the rest of the jobs in $i$ are all good. First, assign $j$ to one machine in $A_1$ that it fits in. Similar to the previous case, we assign the rest of the jobs (that are all good) to different machines in $A_1$. This way, we release machine $i$ and reduce the number of purchased machines by one.
\item 
Find two critical machines $i_1$ and $i_2$. Let $j_1$ and $j_2$ be the only jobs in these two machines that fit in no machine in $A_1$. If both jobs fit and form a feasible assignment in a new machine, we purchase a new machine and assign $j_1$ and $j_2$ to it. Otherwise, we do not change anything and ignore this update step. There are at most $6$ other jobs in these two machines since both are medium machines. In addition, the rest of the jobs are all good. Therefore, similar to the previous two cases, we can assign these jobs to distinct machines in $A_1$ that they fit in. This way, we release machines $i_1$ and $i_2$ and purchase a new machine. So in total, we reduce the number of purchased machines by one. 
\end{itemize}

We are now ready to analyze this $\localsearch$ algorithm that also borrows ideas from $\firstfit$. We next show that the number of purchased machines is at most $2 \opt + O(1)$, i.e., a 2-approximation.

\begin{theorem}\label{offline2}
Algorithm $\localsearch$ terminates after at most $N^3$ operations (where $N$ is the number of jobs), and purchases at most $2 \opt +11$ machines. 
\end{theorem}

The proof can be found in Appendix \ref{proofoffline}.

We conclude this section by comparing our results to the classical (deterministic) bin packing problem. In the classical bin packing problem, there are folklore polynomial time approximation schemes (see Section 10.3 in \cite{CombinatorialAlgorithmsLectureNotes}) that achieve a $(1-\epsilon)$-approximation factor by proposing an offline algorithm based on clustering the jobs into $1/\epsilon^2$ groups, and treating them as equal size jobs. Using dynamic programming techniques, one can solve the simplified problem with $1/\epsilon^2$ different job sizes in time $O(n^{\text{poly}(1/\epsilon)})$. In addition to the inefficient time complexity of these algorithms that make them less appealing for practical purposes, one cannot generalize the same ideas to our setting. The main obstacle is the lack of a total ordering among the different jobs. In the classical bin packing problem, the jobs can be sorted based on their sizes. However, this is not true in our case since the jobs have the two dimensional requirements $\mu_j$ and $b_j$.

\subsection{Alternative constraints}\label{newconst}

Recall that in the (SMBP) problem, we imposed the modified capacity constraint \eqref{general:sub}. Instead, one can consider the following family of constraints, parametrized by $0.5 \leq p \leq 1$:
\begin{align}\label{general:sub:p}
\sum_{j=1}^N \mu_j x_{ij}+ D(\alpha) \Big ( \sum_{j=1}^N b_j x_{ij} \Big )^p \leq V y_i,
\end{align}
Note that this equation is still monotone and submodular in the assignment vector $\mathbf x$, and captures some notion of risk pooling. In particular, the ``safety buffer'' reduces with the number of jobs already assigned to each machine. The motivation behind such a modified capacity constraint lies in the shape that one wishes to impose on the term that captures the uncertain part of the job. In one extreme ($p=1$), we consider that the term that captures the uncertainty is linear and hence, as important as the expectation term. In the other extreme case ($p=0.5$), we consider that the term that captures the uncertainty behaves as a square root term. For a large number of jobs per machine, this is known to be an efficient way of handling uncertainty (similar argument as the central limit theorem). Note also that when $p=0.5$, we are back to equation \eqref{general:sub}, and when $p=1$ we have a commonly used benchmark (see more details in Section \ref{comps}). One can extend our analysis and derive an approximation factor for the online problem as a function of $p$ for any lazy algorithm.

\begin{figure}[h!]
\begin{center}
\includegraphics[width=3.1in]{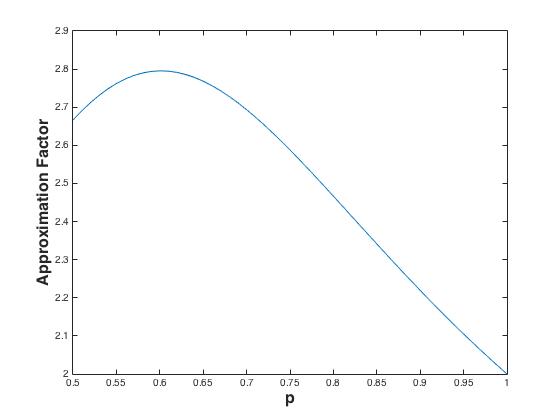}
\end{center}
\caption{Approximation factor $\frac{2} {f(p)}$ as a function of $p$}\label{Approx:SMp}
\end{figure}

\begin{corollary}
Consider the bin packing problem with the modified capacity constraint \eqref{general:sub:p}. Then, any lazy algorithm $\alg$ purchases at most $\frac{2} {f(p)} \opt$ machines, where $\opt$ is the optimum number of machines to serve all jobs and $f(p)$ is given by:
$$
f(p) = 1 - (1-p) p^{\frac{1}{p} - 1}.
$$ 
\end{corollary}
The proof is in a very similar spirit as in Theorem \ref{thm:lazyapp} and is not repeated due to space limitations. Intuitively, we find parametric lower and upper bounds on $\big ( \sum_{j=1}^N b_j x_{ij} \big )^p$ in terms of $\sum_{j=1}^N b_j x_{ij} $. Note that when $p=0.5$, we recover the result of Theorem \ref{thm:lazyapp} (i.e., a 8/3 approximation) and as $p$ increases, the approximation factor converges to 2. In Figure \ref{Approx:SMp}, we plot the approximation factor as a function of $0.5 \leq p \leq 1$. 

Finally, one can also extend our results for the case where the modified capacity constrain is given by: $\sum_{j=1}^N \mu_j x_{ij}+ D(\alpha) \log \Big ( 1 + \sum_{j=1}^N b_j x_{ij} \Big ) \leq V y_i$.
\section{Computational experiments}\label{comps}

In this section, we test and validate the analytical results developed in the paper by solving the (SMBP) problem for different realistic cases,  and investigating the impact on the number of machines required (i.e., the cost). We use realistic workload data inspired by Google Compute Engine, and show how our model and algorithms can be applied in an operational setting.

\subsection{Setting and data}

We use simulated workloads of 1000 jobs (virtual machines) with a realistic VM size distribution (see Table \ref{VMdist}). Typically, the GCE workload is composed of a mix of CPU usages from virtual machines belonging to cloud customers. These jobs can have highly varying workloads, including some large ones and many smaller ones.\footnotemark[8]\footnotetext[8]{The average distribution of workloads we present in Table 1 assumes small percentages of workloads with 32 and 16 cores, and larger percentages of smaller VMs. A ``large'' workload may consist of many VMs belonging to a single customer whose usages may be correlated at the time-scales we are considering, but heuristics ensure these are spread across different hosts to avoid strong correlation of co-scheduled VMs. The workload distributions we are using are representative for some segments of GCE. Unfortunately, we cannot provide the real data due to confidentiality.} More precisely, we assume that each VM arrives to the cloud provider with a requested number of CPU cores, sampled from the distribution presented in Table \ref{VMdist}. 

\begin{table}
\centering
\caption{Example distribution of VM sizes in one Google data center}
\begin{tabular}{c c c c c c c }
\hline
 Number of cores &  1 &  2 &  4 & 8 & 16 & 32 \\
 \hline
\% VMs  & 36.3  & 13.8 & 21.3 & 23.1 & 3.5 & 1.9\\
\hline 
\end{tabular}
\label{VMdist}
\end{table}

In this context, the average utilization is typically low, but in many cases, the utilization can be highly variable over time. Although we decided to keep a similarl VM size distribution as observed in a production data center, we also fitted parametric distributions to roughly match the mean and the variance of the measured usage. This allows us to obtain a parametric model that we could vary for simulation.  We consider two different cases for the actual CPU utilization as a fraction of the requested job size: we either assume that the number of cores used has a Bernoulli distribution, or a truncated Gaussian distribution.  
As discussed in Section \ref{altform}, we assume that each job $j$ has lower and upper utilization bounds, $\underline A_j$ and $\overline A_j$. We sample $\underline A_j$ uniformly in the range $[0.3, 0.6]$, and $\overline A_j$ in the range $[0.7, 1.0]$. In addition, we 
uniformly sample $\mu'_j$ and $\sigma'_j \in [0.1, 0.5]$ for each VM
to serve as the parameters for the truncated Gaussian (not to be confused with its true mean and standard deviation, $\mu_j$ and $\sigma_j$). 
For the Bernoulli case, $\mu_j' = \mu_j$ determines the respective probabilities of the 
realization corresponding to the lower or upper bound (and the unneeded $\sigma'_j$ is ignored). 

For each workload of 1000 VMs generated in this manner, we solve the online version of the (SMBP) problem by implementing the Best-Fit heuristic, using one of the three different variants for the values of $D(\alpha)$ and $b_j$. We solve the problem for various values of $\alpha$ ranging from 0.5 to 0.99999.
More precisely, when a new job arrives, we compute the modified capacity constraint in equation \eqref{general:sub} for each already-purchased machine, and assign the job to the machine with the smallest available capacity that can accommodate it\footnotemark[9]\footnotetext[9]{Note that we clip the value of the constraint at the effective upper bound $\big(\sum_{j} x_{ij} \overline A_j \big)$, to ensure that no trivially feasible assignments are excluded. Otherwise, the Hoeffding's inequality-based constraint may perform slightly worse relative to the policy without over-commitment, if it leaves too much free space on the machines.}. 
If the job does not fit in any of the already purchased machines, the algorithm opens a new machine. We consider the three variations of the (SMBP) discussed earlier:
\begin{itemize}
\item The Gaussian case introduced in \eqref{chance:gauss}, with $b_j = \sigma^2_j$ and $D(\alpha) = \Phi^{-1}(\alpha)$. This is now also an approximation to the chance constrained (BPCC) formulation since the true distributions are truncated Gaussian or Bernoulli.

\item The Hoeffding's inequality approximation introduced in \eqref{HIapprox}, with $b_j = (\overline{A}_j - \underline{A}_j)^2$ and $D(\alpha) = \sqrt{-0.5 \ln(1 - \alpha)}$. 
Note hat the distributionally robust approach with the family of distributions $\mathcal D_2$ is equivalent to this formulation.

\item The distributionally robust approximation with the family of distributions $\mathcal D_1$, with $b_j = \sigma^2_j$ and $D_1(\alpha) = \sqrt{\alpha/(1-\alpha)}$. 
\end{itemize}

\subsection{Linear benchmarks}

We also implement the following four benchmarks which consist of solving the classical
(DBP) problem with specific problem data. First we have:
\begin{itemize}
\item No overcommitment -- This is equivalent to setting $\alpha=1$ in the (SMBP) problem, or 
    solving the (DBP) problem with sizes $\overline A_j$.
\end{itemize}
Three other heuristics are obtained by replacing the square-root term in constraint
\eqref{general:sub} by a linear term, specifically we replace the constraint with:
\begin{align} \label{general:sub:linearized}
\sum_{j=1}^N \mu_j x_{ij}+ D(\alpha) \sum_{j=1}^N \sqrt{b_j x_{ij}} = 
\sum_{j=1}^N \left( \mu_j + D(\alpha) \sqrt{b_j} \right)  x_{ij} \leq V y_i
\end{align}
to obtain:
\begin{itemize}
\item The linear Gaussian heuristic that mimics the Gaussian approximation in \eqref{chance:gauss}. 

\item The linear Hoeffding's heuristic that mimics the Hoeffding's approximation in \eqref{HIapprox}.

\item The linear robust heuristic that mimics the distributionally robust approach with the family of distributions $\mathcal D_1$.
\end{itemize}
Notice that the linearized constraint \eqref{general:sub:linearized} is clearly more restrictive for 
a fixed value of $\alpha$ by concavity of the square root, but we do of course vary the value 
of $\alpha$ in our experiments. We do not expect these benchmarks to outperform
our proposed method since they do not capture the risk-pooling effect from scheduling 
jobs concurrently on the same machine. They do however still reflect
different relative amounts of "padding" or "buffer" above the expected utilization 
of each job allocated due to the usage uncertainty.

The motivation behind the linear benchmarks lies in the fact that the problem is reduced to the standard (DBP) formulation  which admits efficient implementations for the classical heuristics.
For example, the Best-Fit algorithm can
run in time $O(N \log N)$ by maintaining a
list of open machines sorted by the slack left free on each machine
(see \cite{johnson1974nlogn} for details and \emph{linear}-time approximations).
In contrast, our implementation of the Best-Fit heuristic with the
non-linear constraint \eqref{general:sub} takes time $O(N^2)$ since we evaluate the constraint for each machine
when each new job arrives. 
Practically, in cloud VM scheduling systems, this quadratic-time approach may be preferred anyway since it
generalizes straightforwardly to more complex ``scoring'' functions that also take into account additional
factors besides the remaining capacity on a machine, such as multiple resource dimensions, 
performance concerns or correlation between jobs (see, for example, \cite{google2015borg}). In addition, the computational cost could be mitigated by dividing the data center into smaller ``shards'', each consisting of a fraction of the machines, and then trying to assign each 
incoming job only to the machines
in one of the shard. For example, in our experiments we found that there was little
performance advantage in considering sets of more than 1000 jobs at a time.
Nevertheless, our results show that even these linear benchmarks may provide
substantial savings (relative to the no-overcommitment policy) while only requiring very minor changes to classical algorithms: instead of $\overline A_j$, we simply use job sizes defined by $\mu_j$, $b_j$ and $\alpha$.

\subsection{Results and comparisons}

We compare the seven different methods in terms of the number of purchased machines and show that, in most cases, our approach 
significantly reduces the number of machines needed.

We consider two physical machine sizes: 32 cores and 72 cores. As expected, the larger machines achieve a greater benefit from modeling risk-pooling. We draw 50 independent workloads each composed of 1000 VMs as described above. For each workload, we schedule the jobs using the Best-Fit algorithm and report the average number of machines needed across the 50 workloads. Finally, we compute the probability of capacity violation as follows: for each machine used to schedule each of the workloads, we draw 5000 utilization realizations (either from a sum of truncated Gaussian or a sum of Bernoulli distributions), and we count the number of realizations where the total CPU usage of the jobs scheduled on a machine exceeds capacity. 

The sample size was chosen so that our results reflect an effect that is measurable in a typical data center. Since our workloads require on the order of 100 machines each, this corresponds to roughly $50 \times 100 \times 5000=25,000,000$ individual machine-level samples. 
Seen another way, we schedule $50 \times 1000 = 50,000$ jobs and collect 5000 data points from each. 
Assuming a sample is recorded every 10 minutes, say, this corresponds to a few days of traffic even in a small real data center with less than 1000 machines\footnotemark[10]\footnotetext[10]{The exact time needed to collect a comparable data set from a production system depends on the data center size and on the sampling rate, which should be a function of how quickly jobs enter and leave the system, and of how volatile their usages are. 
By sampling independently in our simulations, we are assuming that the measurements from each machine are collected relatively infrequently
 (to limit correlation between successive measurements), and that the workloads are diverse (to limit correlation between measurements from different machines). 
This assumption is increasingly realistic as the size of the data center and the length of time covered increase: in the limit, for a fixed sample size, we would record at most one measurement from each job with a finite lifetime, and it would only be correlated with a small fraction of its peers.}.
The sample turns out to yield very stable measurements, and defining appropriate service level indicators is application-dependent and beyond the scope of this paper, so we do not report confidence intervals or otherwise delve into statistical measurement issues. 
Similarly, capacity planning for smaller data centers may need to adjust measures of demand uncertainty to account for the different scheduling algorithms, but any conclusions are likely specific to the workload and data center, so we do not report on the variability across workloads.

\begin{figure}[h!]
\centering
  \begin{subfigure}[b]{0.43\textwidth}
\includegraphics[width=3.03in]{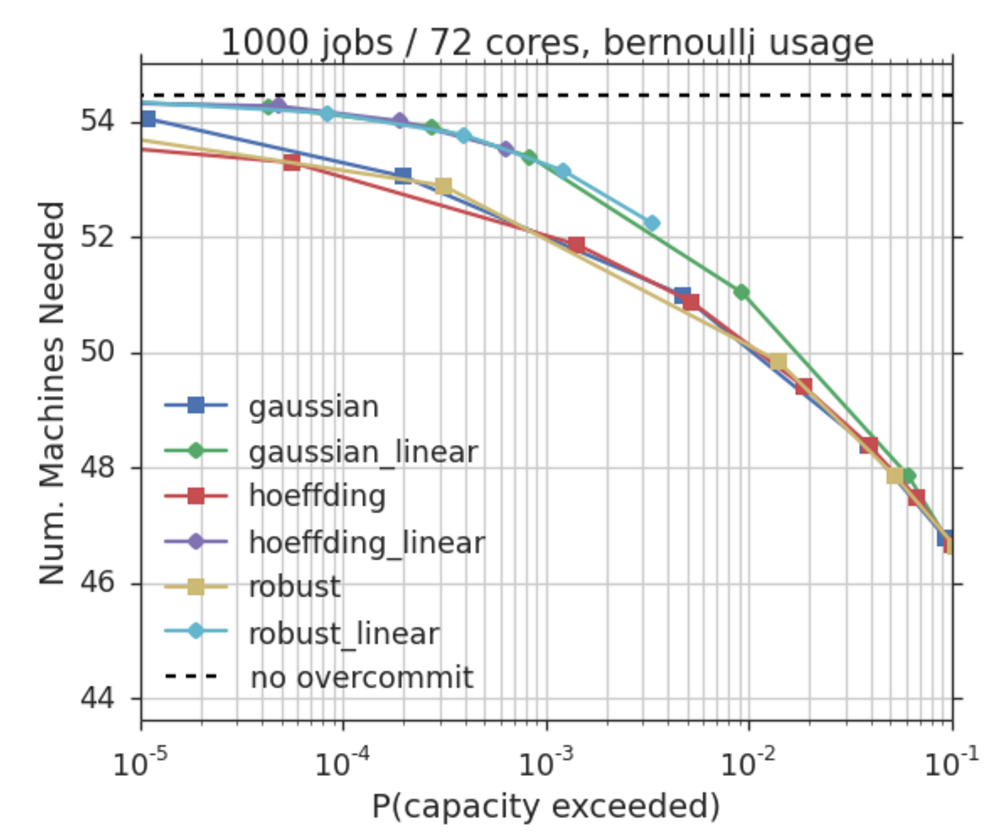}
\caption{Bernoulli usage}\label{72:bern}
  \end{subfigure}
  \hspace{0.04\textwidth}
  \begin{subfigure}[b]{0.43\textwidth}
  \includegraphics[width=2.7in]{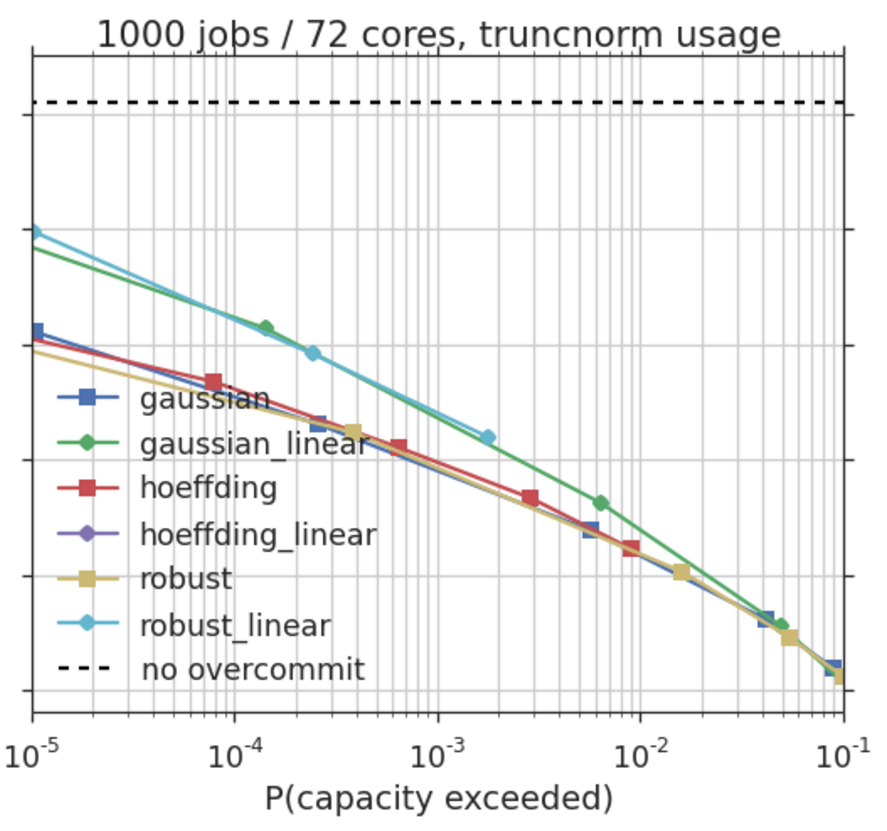}
\caption{Truncated Gaussian usage}\label{72:trunc}
  \end{subfigure}
\caption{Average number of 72-core machines needed to schedule a workload, versus the probability
that any given machine's realized load exceeds capacity}
\label{fig:72cores}
\end{figure}

\begin{figure}[h!]
\begin{center}
\includegraphics[width=5.1in]{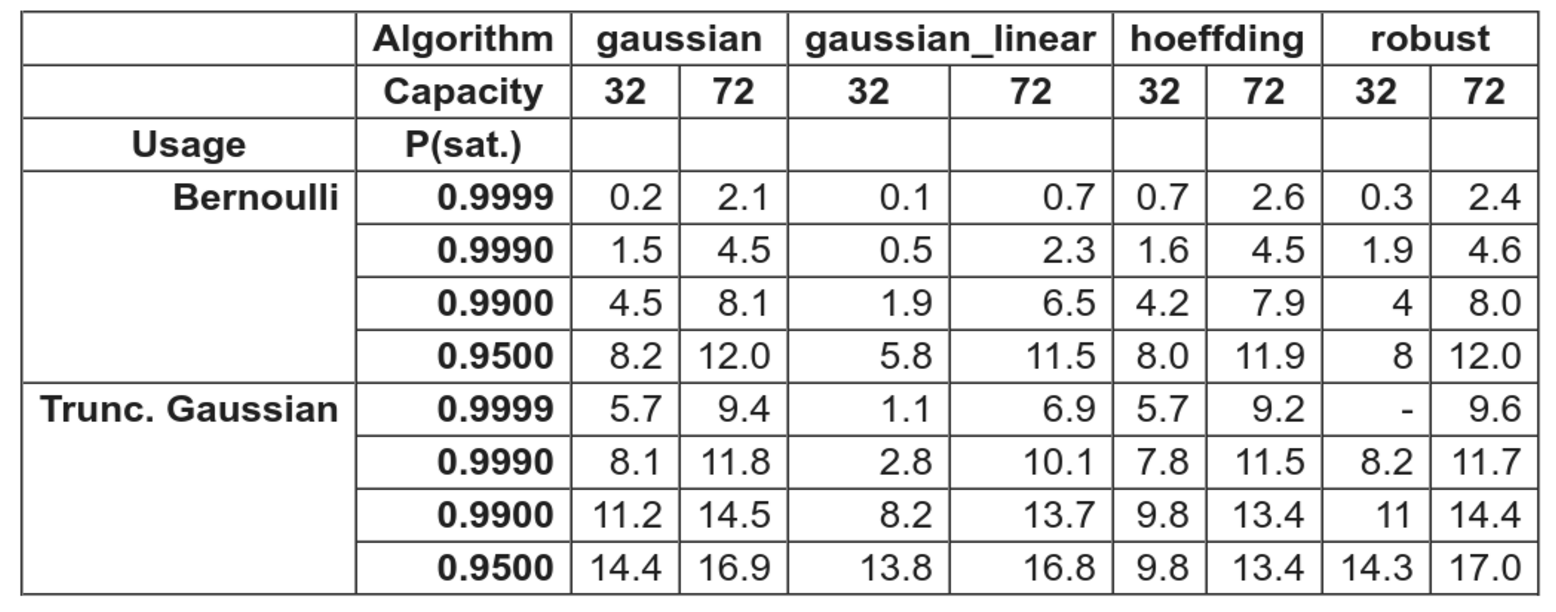}
\end{center}
\caption{Percentage savings due to overcommitment for two CPU usage distributions, 
using the three proposed variants of the chance constraint. 
The linear Gaussian benchmark is shown for comparison 
}\label{Summary:Table}
\end{figure}

In Figure \ref{fig:72cores}, we plot the average number of machines needed as a function of the probability that a given constraint is violated, in the case where the data center is composed of 72 CPU core machines. Each point in the curves corresponds to a different value of the parameter $\alpha$. Without overcommitment, we need an average of over 54 machines in order to serve all the jobs. By allowing a small chance of violation, say a 0.1\% risk (or equivalently, a 99.9\% satisfaction probability), we only need 52 machines for the Bernoulli usage, and 48 machines for the truncated Gaussian usage. If we allow a 1\% chance of violation, we then only need 50 and 46 machines, respectively. The table of Figure \ref{Summary:Table} summarizes the relative savings, which are roughly 4.5\% and 11.5\% with a 0.1\% risk, and roughly 8\% and 14\% with a 1\% risk, for the Bernoulli and truncated Gaussian usages, respectively. In terms of the overcommitment factor defined in Section \ref{overcom}, the reported savings translate directly to the fraction of the final capacity that is due to overcommitment,
$$\frac{B(1) - B(\alpha)}{B(1)} = \frac{OCF(\alpha) - OCF(1)}{OCF(\alpha)}.$$

Figure  \ref{fig:72cores} shows that all three variations of our approach (the Gaussian, Hoeffding's, and the distributionally robust approximations) yield very similar results. This suggests that the results are robust to the method and the parameters. The same is true for the corresponding linear benchmarks, though they perform worse, as expected. 
We remark that although the final performance tradeoff is nearly identical, for a particular value of the $\alpha$ parameter, the achieved violation probabilities vary greatly. For example, with $\alpha=0.9$ and the truncated normal distribution, each constraint was satisfied with probability 0.913 when using the Gaussian approximation, but with much higher probabilities 0.9972 and 0.9998 for the Hoeffding and robust approximations, respectively. This is expected, since the latter two are relatively loose upper bounds for a truncated normal distribution, whereas the distributions $N(\mu_j, \sigma_j)$ are close approximations to the truncated Gaussian with parameters $\mu_j'$ and $\sigma_j'$.
 (This is especially true for their respective sums.)
Practically, the normal approximation is likely to be the easiest to calibrate and understand in cases where the theoretical guarantees of the other two approaches are not needed, since it would be nearly exact for normally-distributed usages.

\begin{figure}[h!]
\centering
  \begin{subfigure}[b]{0.43\textwidth}
\includegraphics[width=3.03in]{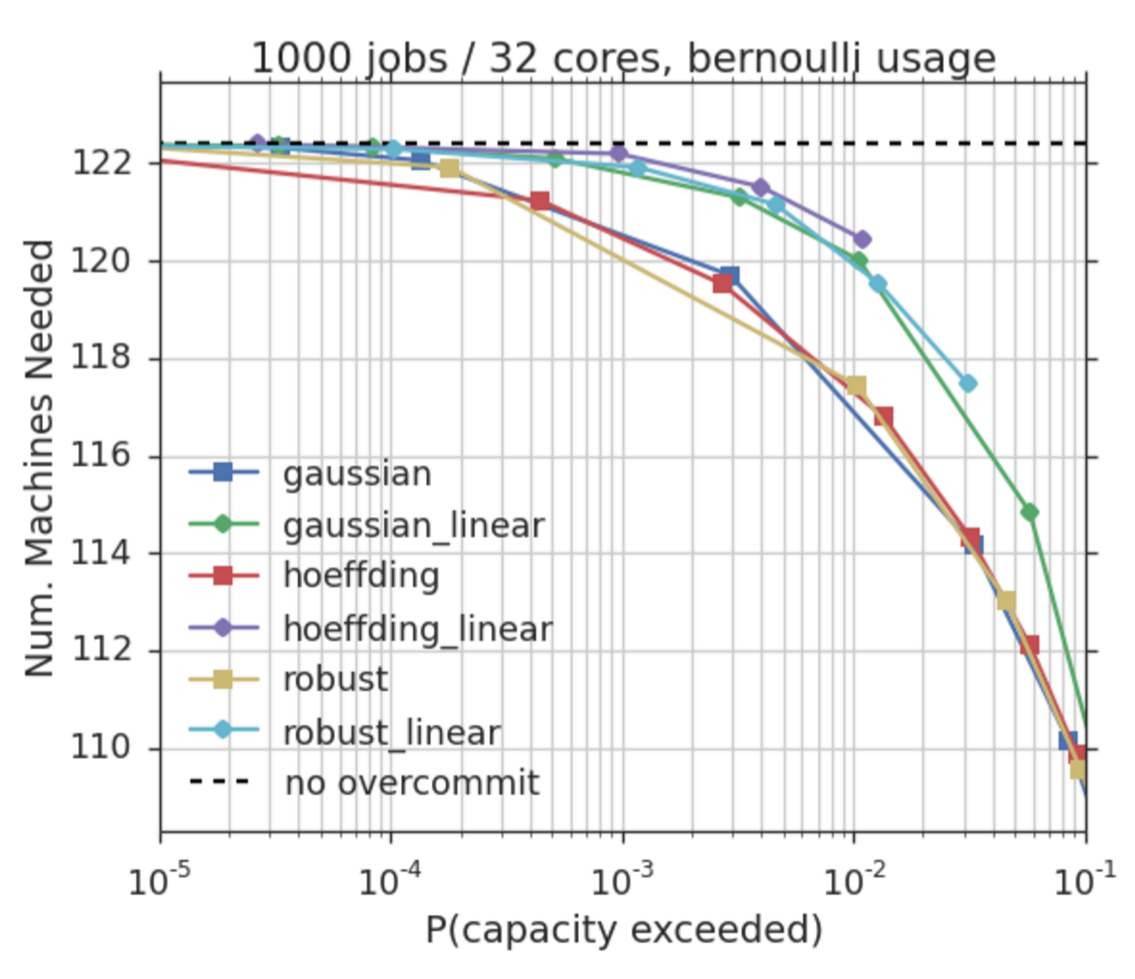}
\caption{Bernoulli usage}\label{72:bern}
  \end{subfigure}
  \hspace{0.04\textwidth}
  \begin{subfigure}[b]{0.43\textwidth}
  \includegraphics[width=2.89in]{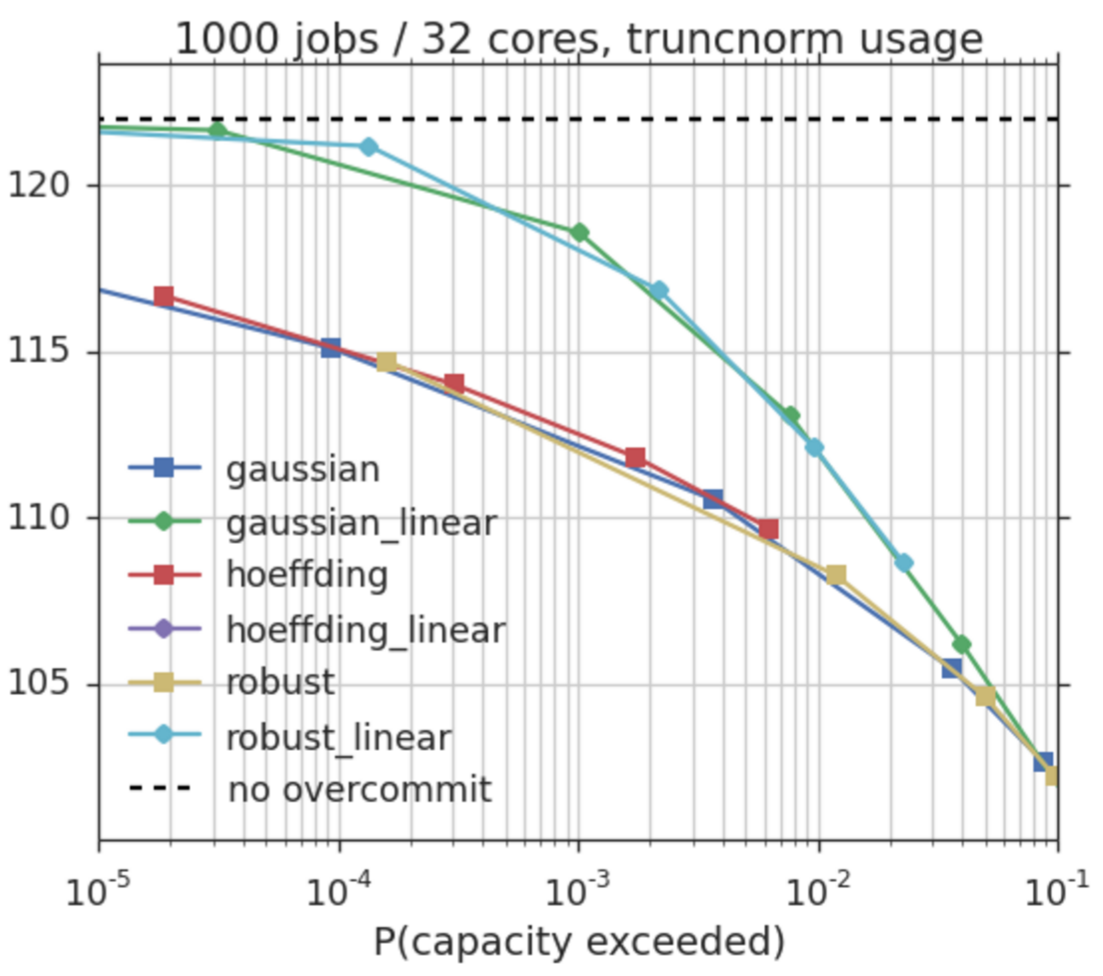}
\caption{Truncated Gaussian usage}\label{32:trunc}
  \end{subfigure}
\caption{Results for 32 core machines}
\label{fig:32cores}
\end{figure}

In Figure \ref{fig:32cores}, we repeat the same tests for smaller machines having only 32 physical CPU cores. 
The smaller machines are more difficult to overcommit since there is a smaller risk-pooling opportunity, as can be seen
by comparing the columns of Table \ref{Summary:Table}.
The three variations of our approach still yield similar and significant savings, but now they substantially 
outperform the linear benchmarks: the cost reduction is at least double with all but the largest
values of $\alpha$. We highlight that with the  ``better behaved'' truncated Gaussian usage, we still obtain
a 5\% cost savings at 0.01\% risk, whereas the linear benchmarks barely improve over the no-overcommit case.


As mentioned in Section \ref{chanceconst}, the value of $\alpha$ should be calibrated so as to yield an acceptable risk level given the data center, the workload and the resource in question. 
Any data center has a baseline risk due to machine (or power) failure, say, 
and a temporary CPU shortage is usually much less severe relative to such a failure.
On the other hand, causing a VM to crash because of a memory shortage can be as bad as a machine failure from the customer's point of view. 
Ultimately, the risk tolerance will be driven by technological factors, such as the ability to 
migrate VMs or swap memory while maintaining an acceptable performance.

\subsection{Impact}

We conclude that our approach allows a substantial cost reduction for realistic workloads. More precisely, we draw the following four conclusions.
\begin{itemize}
\item \textbf{Easy to implement:} Our approach is nearly as simple to implement as classical bin packing heuristics. In addition, it works naturally online and in real-time, and can be easily incorporated to existing scheduling algorithms.

\item \textbf{Robustness:} The three variations we proposed yield very similar results. This suggests that our approach is robust to the type of approximation. In particular, the uncertain term $b_j$ and the risk coefficient $D(\alpha)$ do not have a strong impact on the results. It also suggests that the method is robust to estimation errors in the measures of variability that define $b_j$.

\item \textbf{Significant cost reduction:} With modern 72-core machines, our approach allows a 8-14\% cost savings relative to the no overcommitment policy. This is achieved by considering a manageable risk level of 1\%, which is comparable to other sources of risk that are not controllable (e.g., physical failures and regular maintenance operations).

\item \textbf{Outperforming the benchmarks:} Our proposals show a consistent marked improvement over three different ``linear'' benchmarks that reduce to directly apply the classical Best-Fit heuristic. The difference is most substantial in cases where the machines are small relative to the jobs they must contain, which is intuitively more challenging. Although our approach does not run in $O(n \log n)$ time, "sharding" and (potentially) parallelization mitigate any such concerns in practice.

\end{itemize}

\section{Conclusion}\label{concl}

In this paper, we formulated and practically solved the bin-packing problem with overcommitment. In particular, we focused on a cloud computing provider that is willing to overcommit when allocating capacity to virtual machines in a data center. We modeled the problem as bin packing with chance constraints, where the objective is to minimize the number of purchased machines, while satisfying the physical capacity constraints of each machine with a very high probability. We first showed that this problem is closely related to an alternative formulation that we call the SMBP (Submodular Bin Packing) problem. Specifically, the two problems are equivalent under the assumption of independent Gaussian job sizes, or when the job size distribution belongs to the distributionally robust family with a given mean and (diagonal) covariance matrix. In addition, the bin packing problem with chance constraints can be approximated by the SMBP for distributions with bounded supports.

We first showed that for the bin packing problem with general monotone submodular capacity constraints, it is impossible to find a solution within any reasonable factor from optimal. 
We then developed simple algorithms that achieve solutions within constant factors from optimal for the SMBP problem. We showed that any lazy algorithm is 8/3 competitive, and that the First-Fit heuristic is 9/4 competitive. Since the First-Fit and Best-Fit algorithms are easy to implement and well understood in practice, this provides an attractive option from an implementation perspective. Second, we proposed an algorithm for the offline version of the problem, and showed that it guarantees a 2-approximation. Then, we used our model and algorithms in order to draw several useful insights on how to schedule jobs to machines, and on the right way to overcommit. We convey that our method captures the risk pooling effect, as the ``safety buffer'' needed for each job decreases with the number of jobs already assigned to the same machine. Moreover, our approach translates to a transparent and meaningful recipe on how to assign jobs to machines by naturally clustering similar jobs in terms of statistical information. Namely, jobs with a similar ratio $b / \mu$ (the uncertain term divided by the expectation) should be assigned to the same machine.

Finally, we demonstrated the benefit of overcommitting and applied our approach to realistic workload data inspired by Google Compute Engine. We showed that our methods are (i) easy to implement; (ii) robust to the parameters; and (iii) significantly reduce the cost (1.5-17\% depending on the setting and the size of the physical machines in the data center).

\ACKNOWLEDGMENT{We would like to thank the Google Cloud Analytics team for helpful discussions and feedback. The first author would like to thank Google Research as this work would not have been possible without a one year postdoc at Google NYC during the year 2015-2016. The authors would also like to thank Lennart Baardman, Arthur Flajolet and Balasubramanian Sivan for their valuable feedback that has helped us improve the paper.}

\bibliographystyle{informs2014}
\bibliography{biblio_GCEO}

\begin{thebibliography}{38}
\providecommand{\natexlab}[1]{#1}
\providecommand{\url}[1]{\texttt{#1}}
\providecommand{\urlprefix}{URL }

\bibitem[{Abdelaziz et~al.(2007)Abdelaziz, Aouni, \protect\BIBand{}
  El~Fayedh}]{abdelaziz2007multi}
Abdelaziz FB, Aouni B, El~Fayedh R (2007) Multi-objective stochastic
  programming for portfolio selection. \emph{European Journal of Operational
  Research} 177(3):1811--1823.

\bibitem[{Alan~Roytman(2013)}]{microsoft2013performanceaware}
Alan~Roytman SGJLSN Aman~Kansal (2013) Algorithm design for performance aware
  vm consolidation. Technical report,
  \urlprefix\url{https://www.microsoft.com/en-us/research/publication/algorithm-design-for-performance-aware-vm-consolidation/}.

\bibitem[{Albers \protect\BIBand{}
  Souza(2011)}]{CombinatorialAlgorithmsLectureNotes}
Albers S, Souza A (2011) Combinatorial algorithms lecture notes: Bin packing.
  \urlprefix\url{https://www2.informatik.hu-berlin.de/alcox/lehre/lvws1011/coalg/bin_packing.pdf}.

\bibitem[{Anily et~al.(1994)Anily, Bramel, \protect\BIBand{}
  Simchi-Levi}]{anily1994worst}
Anily S, Bramel J, Simchi-Levi D (1994) Worst-case analysis of heuristics for
  the bin packing problem with general cost structures. \emph{Operations
  research} 42(2):287--298.

\bibitem[{Bays(1977)}]{bays1977comparison}
Bays C (1977) A comparison of next-fit, first-fit, and best-fit.
  \emph{Communications of the ACM} 20(3):191--192.

\bibitem[{Bertsimas \protect\BIBand{} Popescu(2005)}]{bertsimas2005optimal}
Bertsimas D, Popescu I (2005) Optimal inequalities in probability theory: A
  convex optimization approach. \emph{SIAM Journal on Optimization}
  15(3):780--804.

\bibitem[{Calafiore \protect\BIBand{}
  El~Ghaoui(2006)}]{calafiore2006distributionally}
Calafiore GC, El~Ghaoui L (2006) On distributionally robust chance-constrained
  linear programs. \emph{Journal of Optimization Theory and Applications}
  130(1):1--22.

\bibitem[{Charnes \protect\BIBand{} Cooper(1963)}]{charnes1963deterministic}
Charnes A, Cooper WW (1963) Deterministic equivalents for optimizing and
  satisfying under chance constraints. \emph{Operations research} 11(1):18--39.

\bibitem[{Coffman et~al.(1980)Coffman, So, Hofri, \protect\BIBand{}
  Yao}]{coffman1980stochastic}
Coffman EG, So K, Hofri M, Yao A (1980) A stochastic model of bin-packing.
  \emph{Information and Control} 44(2):105--115.

\bibitem[{Coffman~Jr et~al.(1996)Coffman~Jr, Garey, \protect\BIBand{}
  Johnson}]{coffman1996approximation}
Coffman~Jr EG, Garey MR, Johnson DS (1996) Approximation algorithms for bin
  packing: a survey. \emph{Approximation algorithms for NP-hard problems},
  46--93 (PWS Publishing Co.).

\bibitem[{Csirik et~al.(2006)Csirik, Johnson, Kenyon, Orlin, Shor,
  \protect\BIBand{} Weber}]{csirik2006sum}
Csirik J, Johnson DS, Kenyon C, Orlin JB, Shor PW, Weber RR (2006) On the
  sum-of-squares algorithm for bin packing. \emph{Journal of the ACM (JACM)}
  53(1):1--65.

\bibitem[{de~La~Vega \protect\BIBand{} Lueker(1981)}]{de1981bin}
de~La~Vega WF, Lueker GS (1981) Bin packing can be solved within 1+
  $\varepsilon$ in linear time. \emph{Combinatorica} 1(4):349--355.

\bibitem[{Delage \protect\BIBand{} Ye(2010)}]{delage2010distributionally}
Delage E, Ye Y (2010) Distributionally robust optimization under moment
  uncertainty with application to data-driven problems. \emph{Operations
  research} 58(3):595--612.

\bibitem[{Delorme et~al.(2016)Delorme, Iori, \protect\BIBand{}
  Martello}]{delorme2016exactcsp}
Delorme M, Iori M, Martello S (2016) Bin packing and cutting stock problems:
  Mathematical models and exact algorithms. \emph{European Journal of
  Operational Research} 255(1):1 -- 20, ISSN 0377-2217,
  \urlprefix\url{http://dx.doi.org/http://dx.doi.org/10.1016/j.ejor.2016.04.030}.

\bibitem[{Dinh et~al.(2013)Dinh, Lee, Niyato, \protect\BIBand{}
  Wang}]{dinh2013survey}
Dinh HT, Lee C, Niyato D, Wang P (2013) A survey of mobile cloud computing:
  architecture, applications, and approaches. \emph{Wireless communications and
  mobile computing} 13(18):1587--1611.

\bibitem[{D{\'o}sa(2007)}]{dosa2007tight}
D{\'o}sa G (2007) The tight bound of first fit decreasing bin-packing algorithm
  is $ffd \leq 11/9 opt + 6/9$. \emph{Combinatorics, Algorithms, Probabilistic
  and Experimental Methodologies}, 1--11 (Springer).

\bibitem[{D{\'o}sa \protect\BIBand{} Sgall(2013)}]{dosa2013first}
D{\'o}sa G, Sgall J (2013) First fit bin packing: A tight analysis.
  \emph{LIPIcs-Leibniz International Proceedings in Informatics}, volume~20
  (Schloss Dagstuhl-Leibniz-Zentrum fuer Informatik).

\bibitem[{Fox et~al.(2009)Fox, Griffith, Joseph, Katz, Konwinski, Lee,
  Patterson, Rabkin, \protect\BIBand{} Stoica}]{fox2009above}
Fox A, Griffith R, Joseph A, Katz R, Konwinski A, Lee G, Patterson D, Rabkin A,
  Stoica I (2009) Above the clouds: A berkeley view of cloud computing.
  \emph{Dept. Electrical Eng. and Comput. Sciences, University of California,
  Berkeley, Rep. UCB/EECS} 28(13):2009.

\bibitem[{Ghaoui et~al.(2003)Ghaoui, Oks, \protect\BIBand{}
  Oustry}]{ghaoui2003worst}
Ghaoui LE, Oks M, Oustry F (2003) Worst-case value-at-risk and robust portfolio
  optimization: A conic programming approach. \emph{Operations Research}
  51(4):543--556.

\bibitem[{Gilmore \protect\BIBand{} Gomory(1961)}]{gilmore1961linear}
Gilmore PC, Gomory RE (1961) A linear programming approach to the cutting-stock
  problem. \emph{Operations research} 9(6):849--859.

\bibitem[{Goemans et~al.(2009)Goemans, Harvey, Iwata, \protect\BIBand{}
  Mirrokni}]{goemans2009approximating}
Goemans MX, Harvey NJ, Iwata S, Mirrokni V (2009) Approximating submodular
  functions everywhere. \emph{Proceedings of the twentieth Annual ACM-SIAM
  Symposium on Discrete Algorithms}, 535--544 (Society for Industrial and
  Applied Mathematics).

\bibitem[{Gupta \protect\BIBand{} Radovanovic(2012)}]{gupta2012online}
Gupta V, Radovanovic A (2012) Online stochastic bin packing. \emph{arXiv
  preprint arXiv:1211.2687} .

\bibitem[{Johnson(1974)}]{johnson1974nlogn}
Johnson DS (1974) Fast algorithms for bin packing. \emph{Journal of Computer
  and System Sciences} 8(3):272 -- 314, ISSN 0022-0000,
  \urlprefix\url{http://dx.doi.org/http://dx.doi.org/10.1016/S0022-0000(74)80026-7}.

\bibitem[{Karaesmen \protect\BIBand{}
  Van~Ryzin(2004)}]{karaesmen2004overbooking}
Karaesmen I, Van~Ryzin G (2004) Overbooking with substitutable inventory
  classes. \emph{Operations Research} 52(1):83--104.

\bibitem[{Keller et~al.(2012)Keller, Tighe, Lutfiyya, \protect\BIBand{}
  Bauer}]{keller2012analysis}
Keller G, Tighe M, Lutfiyya H, Bauer M (2012) An analysis of first fit
  heuristics for the virtual machine relocation problem. \emph{Network and
  service management (cnsm), 2012 8th international conference and 2012
  workshop on systems virtualiztion management (svm)}, 406--413 (IEEE).

\bibitem[{Kenyon et~al.(1996)}]{kenyon1996best}
Kenyon C, et~al. (1996) Best-fit bin-packing with random order. \emph{SODA},
  volume~96, 359--364.

\bibitem[{Lueker(1983)}]{lueker1983bin}
Lueker GS (1983) Bin packing with items uniformly distributed over intervals
  [a, b]. \emph{Foundations of Computer Science, 1983., 24th Annual Symposium
  on}, 289--297 (IEEE).

\bibitem[{Nemirovski \protect\BIBand{} Shapiro(2006)}]{nemirovski2006convex}
Nemirovski A, Shapiro A (2006) Convex approximations of chance constrained
  programs. \emph{SIAM Journal on Optimization} 17(4):969--996.

\bibitem[{Pisinger \protect\BIBand{} Sigurd(2005)}]{pisinger2005two}
Pisinger D, Sigurd M (2005) The two-dimensional bin packing problem with
  variable bin sizes and costs. \emph{Discrete Optimization} 2(2):154--167.

\bibitem[{Rina~Panigrahy(2011)}]{microsoft2011validating}
Rina~Panigrahy KTUWRR Vijayan~Prabhakaran (2011) Validating heuristics for
  virtual machines consolidation. Technical report,
  \urlprefix\url{https://www.microsoft.com/en-us/research/publication/validating-heuristics-for-virtual-machines-consolidation/}.

\bibitem[{Rothstein(1971)}]{rothstein1971airline}
Rothstein M (1971) An airline overbooking model. \emph{Transportation Science}
  5(2):180--192.

\bibitem[{Rothstein(1985)}]{rothstein1985or}
Rothstein M (1985) Or forum -- or and the airline overbooking problem.
  \emph{Operations Research} 33(2):237--248.

\bibitem[{Sindelar et~al.(2011)Sindelar, Sitaraman, \protect\BIBand{}
  Shenoy}]{google2011vmpacking}
Sindelar M, Sitaraman R, Shenoy P (2011) Sharing-aware algorithms for virtual
  machine colocation. \emph{Proceedings of the 23rd ACM symposium on
  Parallelism in algorithms and architectures}, 367--378 (New York, NY, USA).

\bibitem[{Stolyar \protect\BIBand{} Zhong(2015)}]{stolyar2015asymptotic}
Stolyar AL, Zhong Y (2015) Asymptotic optimality of a greedy randomized
  algorithm in a large-scale service system with general packing constraints.
  \emph{Queueing Systems} 79(2):117--143.

\bibitem[{Subramanian et~al.(1999)Subramanian, Stidham~Jr, \protect\BIBand{}
  Lautenbacher}]{subramanian1999airline}
Subramanian J, Stidham~Jr S, Lautenbacher CJ (1999) Airline yield management
  with overbooking, cancellations, and no-shows. \emph{Transportation Science}
  33(2):147--167.

\bibitem[{Svitkina \protect\BIBand{} Fleischer(2011)}]{SvitkinaFleischer}
Svitkina Z, Fleischer L (2011) Submodular approximation: Sampling-based
  algorithms and lower bounds. \emph{SIAM Journal on Computing}
  40(6):1715--1737.

\bibitem[{Verma et~al.(2015)Verma, Pedrosa, Korupolu, Oppenheimer, Tune,
  \protect\BIBand{} Wilkes}]{google2015borg}
Verma A, Pedrosa L, Korupolu MR, Oppenheimer D, Tune E, Wilkes J (2015)
  Large-scale cluster management at {Google} with {Borg}. \emph{Proceedings of
  the European Conference on Computer Systems (EuroSys)} (Bordeaux, France).

\bibitem[{Weatherford \protect\BIBand{} Bodily(1992)}]{weatherford1992taxonomy}
Weatherford LR, Bodily SE (1992) A taxonomy and research overview of
  perishable-asset revenue management: yield management, overbooking, and
  pricing. \emph{Operations Research} 40(5):831--844.

\end{thebibliography}

\newpage
\begin{APPENDICES}

\section{Details for Section \ref{naive}}\label{append:naive}

\subsection*{IP formulation}
By taking the square on both sides of the submodular capacity constraint \eqref{general:sub}, we obtain:
\begin{align*} 
V^2 y_i + \Big ( \sum_{j=1}^N \mu_j x_{ij} \Big)^2 - 2 V y_i \sum_{j=1}^N \mu_j x_{ij} \geq D (\alpha)^2 \cdot \sum_{j=1}^N b_j x_{ij}.
\end{align*}
Note that since $y_i$ is binary, we have $y_i^2 = y_i$. We next look at the term: $y_i \sum_{j=1}^N \mu_j x_{ij}$. One can linearize this term by using one of the following two methods. 

\begin{enumerate}
\item Since $y_i = 1$ if and only if at least one $x_{ij} =1$, we have the constraint: $\sum_{j=1}^N  x_{ij} \leq M y_i$, for a large positive number $M$ (actually, one can take $M =N$). Consequently, one can remove the $y_i$ in the above term.

\item One can define a new variable  $t_{ij}  \triangleq y_i x_{ij}$ and add the four following constraints:
\begin{align*}
t_{ij} \leq y_i; ~~ t_{ij} \leq x_{ij}; ~~t_{ij} \geq 0; ~~ t_{ij} \geq x_{ij} + y_i - 1.
\end{align*}
\end{enumerate}
Next, we look at the term: $\Big ( \sum_{j=1}^N \mu_j x_{ij} \Big)^2 $. Since $x_{ij}^2 = x_{ij}$, we remain only with the terms $x_{ij} \cdot x_{ik}$ for $k>j$. One can now define a new variable for each such term, i.e., $z_{ijk}  \triangleq x_{ij} \cdot x_{ik}$ with the four constraints as before:
\begin{align*}
z_{ijk} \leq x_{ij} ; ~~ z_{ijk} \leq x_{ik}; ~~z_{ijk} \geq 0; ~~ z_{ijk}  \geq x_{ij} +x_{ik}  - 1.
\end{align*}
The resulting formulation is a linear integer program. Note that the decision variables $t_{ij} $ and $z_{ijk} $ are continuous, and only $x_{ij}$ and $y_i$ are binary. 

\section{Proof of Theorem \ref{general:submod}}\label{submod:proof}
\proof{Proof.}
In this proof, we make use of the submodular functions defined by \cite{SvitkinaFleischer} for load balancing problems.  
Denote the jobs by $1, 2, \cdots, N$, and for every subset of jobs $S \subseteq [N]$, let $f(S)$ be the cost of the set $S$ (i.e., the capacity cost induced by the function $f$). We use two submodular functions $f$ and $f'$ (defined formally next) which are proved to be indistinguishable with a polynomial number of value oracle queries (see Lemma 5.1 of \cite{SvitkinaFleischer}). Let  denote $x = \ln(N)$. Note that \cite{SvitkinaFleischer} require $x$ to be any parameter such that $x^2$ dominates $\ln(N)$ asymptotically and hence, includes the special case we are considering here.
  Define $m_0 = \frac{5\sqrt{N}}{x}$, $\alpha_0 = \frac{N}{m_0}$, and $\beta_0 = \frac{x^2}{5}$. We choose $N$ such that $m_0$ takes an integer value. Define $f(S)$ to be $\min\{|S|, \alpha_0\}$, and $f'(S)$ to be $\min\{\sum_{i} \min\{\beta_0, |S \cap V_i|\}, \alpha_0\}$ where $\{V_i\}_{i=1}^{m_0}$ is a random partitioning of $[N]$ into $m_0$ equal sized parts. 
Note that by definition, both set functions $f$ and $f'$ are monotone and submodular. 

As we mentioned, it is proved in \cite{SvitkinaFleischer} that the submodular functions $f$ and $f'$ cannot be distinguished from each other with a polynomial number of value oracle queries   with high probability. 
We construct two instances of the bin packing problem with monotone submodular capacity constraints by using  $f$ and $f'$ as follows. 
In both instances, the capacity of each machine is set to $\beta_0$. 

In the first instance, a set $S$ is feasible (i.e., we can schedule all its jobs in a machine) if and only if $f(S) \leq \beta_0$. 
By definition, $f(S)$ is greater than $\beta_0$ if $|S|$ is greater than $\beta_0$. Therefore, any feasible set $S$ in this first instance consists of at most $\beta_0$ jobs. Consequently, in the first instance, any feasible assignment of jobs to machines requires at least $\frac{N}{\beta_0}$ machines. 

We define the second instance of the bin packing problem based on the submodular function $f'$. 
A set $S$ is feasible in the second instance,  if and only if $f'(S) \leq \beta_0$. 
Since $f'(V_j)$ is at most $\beta_0$ for each $1 \leq j \leq m_0$, each set $V_j$ is a feasible set in this second instance. Therefore, we can assign each $V_j$ to a separate machine to process all jobs, and consequently, $m_0$ machines suffice to do all the tasks in the second instance. We note that with our parameter setting, $m_0$ is much smaller that $\frac{N}{\beta_0}$. We then conclude that the optimum solutions of these two instances differ significantly. 

We next prove the claim of Theorem \ref{general:submod} by using a contradiction argument. Assume that there exists a polynomial time algorithm \textsc{ALG} for the bin packing problem with monotone submodular capacity constraints with an approximation factor better than $\frac{\sqrt{N}}{\ln(N)}$. We next prove that by using \textsc{ALG}, we can distinguish between the two set functions $f$ and $f'$ with a polynomial number of value oracles, which contradicts the result of \cite{SvitkinaFleischer}. The task of distinguishing between the two functions $f$ and $f'$ can be formalized as follows. We have value oracle access to a set function $g$, and we know that $g$ is either the same as $f$ or the same as $f'$. The goal is to find out whether $g=f$ or $g=f'$ using a polynomial number of value oracle queries. We construct a bin packing instance with $N$ jobs, capacity constraints $g$, and ask the algorithm \textsc{ALG} to solve this instance. If \textsc{ALG} uses less than $\frac{N}{\beta_0}$ machines to process all jobs, we can say that $g$ is the same as $f'$ (since with capacity constraints $f$, there does not exist a feasible assignment of all jobs to less than $\frac{N}{\beta_0}$ machines). 
On the other hand, if  \textsc{ALG} uses at least  $\frac{N}{\beta_0}$ machines, we can say that $g$ is equal to $f$. This follows from the fact that if $g$ was equal to $f'$, the optimum number of machines would have been at most $m_0$. Since \textsc{ALG} has an approximation factor better than $\frac{\sqrt{N}}{\ln(N)}$, the number of machines used by \textsc{ALG} should have been less than $m_0 \times \frac{\sqrt{N}}{\ln(N)} = \frac{5\sqrt{N}}{x} \times \frac{\sqrt{N}}{\ln(N)} = \frac{N}{\beta_0}$. Therefore, using at least  $\frac{N}{\beta_0}$ machines by \textsc{ALG} is a sufficient indicator of $g$ being the same as $f$. This argument implies that an algorithm with an approximation factor better than $\frac{\sqrt{N}}{\ln(N)}$ for the bin packing problem with monotone submodular constraints yields a way of distinguishing between $f$ and $f'$ with a polynomial number of value oracle queries (since \textsc{ALG} is a polynomial time algorithm), which contradicts the result of \cite{SvitkinaFleischer}.
    $\Box$ 
    
\endproof

\section{Details related to Observation \ref{CSP:obs}}\label{CSP:details}

For ease of exposition, we first address the case with two job classes. Classes 1 and 2 have parameters $(\mu_1, b_1)$ and $(\mu_2, b_2)$ respectively. For example, an interesting special case is when one class of jobs is more predictable relative to the other (i.e., $\mu_1 = \mu_2 = \mu$, $b_2 = b$ and $b_1 = 0$). In practice, very often, one class of jobs has low variability (i.e., close to deterministic), whereas the other class is more volatile. For example, class 1 can represent loyal recurring customers, whereas class 2 corresponds to new customers.

We assume that we need to decide the number of machines to purchase, as well as how many jobs of types 1 and 2 to assign to each machine. Our goal is to find the right mix of jobs of classes 1 and 2 to assign to each machine (note that this proportion can be different for each machine). Consider a given machine $i$ and denote by $n_1$ and $n_2$ the number of jobs of classes 1 and 2 that we assign to this machine. We would like to ensure that the chance constraint is satisfied in each machine with the given parameter $\alpha$. Assuming that $V > n_1 \mu_1 + n_2 \mu_2 $, we obtain:
\begin{align}\label{alpha:N:2}
\frac{[ V - n_1 \mu_1 - n_2 \mu_2 ]^2} {n_1 b_1 + n_2 b_2}  = D(\alpha)^2.
\end{align}
For a given  $\alpha$, one can find the value of $n_1$ as a function of $n_2$ that satisfies equation \eqref{alpha:N:2}:
\begin{align}\label{N:alpha:2}
n_1(n_2) = \frac{V - n_2 \mu_2} {\mu_1} + \frac{1} {2 \mu_1^2} \big [ b_1D(\alpha)^2  - \sqrt{ b_1^2 D(\alpha)^4 + 4 b_1 D(\alpha)^2 (V - n_2 \mu_2) \mu_1 + 4 \mu_1^2 n_2 b_2 D(\alpha)^2} \big].
\end{align}

As we discussed, an interesting special case is when both classes of jobs have the same expectation, i.e., $\mu_1 = \mu_2 = \mu$ but one type of jobs is much more predictable (i.e., smaller range or variance). In the extreme case, one can assume that class 1 jobs are deterministic, (i.e., $b_1 = 0$). In this case, equation \eqref{N:alpha:2} becomes:
\begin{align}\label{N:alpha:2:extreme}
n_1(n_2) = \frac{V - n_2 \mu} {\mu} - \frac{\sqrt{2 n_2 b_2 B}} {2 \mu}.
\end{align}
Alternatively, by directly looking at the modified capacity constraint \eqref{general:sub} for this special case, we obtain:
\begin{align}\label{special:case}
V = (n_1 + n_2) \mu + D(\alpha)  \sqrt{n_2 b_2}.
\end{align}
Equation \eqref{special:case} can be interpreted as follows. Any additional job of type 1 takes $\mu$ from the capacity budget $V$, whereas any additional job of type 2 is more costly. The extra cost depends on both the uncertainty of the job (through $b_2$) and the overcommitment policy (through $ D(\alpha)$). The higher is one of these two factors, the larger is the capacity we should plan for jobs of type 2 (i.e., ``safety buffer''). Note that the submodular nature of constraint \eqref{general:sub} implies that this marginal extra cost decreases with the number of jobs $n_2$. In other words, when $n_2$ becomes large, each additional job of type 2 will converge to take a capacity of $\mu$, as the central limit theorem applies. More generally,  by taking the derivative of the above expression, any additional job of type 2 will take $\mu + 0.5 D(\alpha) \sqrt {b_2}  / \sqrt{n_2}$ (where $n_2$ here represents how many jobs of type 2 are already assigned to this machine).

In Figure \ref{alpha:N:2classes}, we plot equation \eqref{N:alpha:2:extreme} for a specific instance with $V=30$ and different values of $\alpha$. As expected, if $n_2=0$, we can schedule $n_1=50$ jobs of class 1 to reach exactly the capacity $V$, no matter what is the value of $\alpha$. On the other hand, for $\alpha=0.99$, if $n_1=0$, we can schedule $n_2=38$ jobs of class 2. As the value of $n_2$ increases, the optimal value of $n_1$ decreases. For a given value of $\alpha$, any point $(n_1, n_2)$ on the curve (or below) guarantees the feasibility of the chance constraint. The interesting insight is to characterize the proportion of jobs of classes 1 and 2 per machine. For example, if we want to impose $n_1=n_2$ in each machine, what is the optimal value for a given $\alpha$? In our example, when $\alpha=0.99$, we can schedule $n_1 = n_2 = 21$ jobs. If we compare to the case without overcommitment (i.e., $\alpha=1$), we can schedule 18 jobs from each class. Therefore, we obtain an improvement of 16.67\%. More generally, if the cost (or priority) of certain jobs is higher, we can design an optimal ratio per machine so that it still guarantees to satisfy the chance constraint. To summarize, for the case when $V=30$, $\mu =0.65$, $\overline A = 1$, $\underline A =0.3$ and $\alpha = 0.99$, one can schedule either 50 jobs of class 1 or 38 jobs of class 2 or any combination of both classes according to equation \eqref{N:alpha:2:extreme}. In other words, we have many different ways of bin packing jobs of classes 1 and 2 to each machine. 

\begin{figure}[h!]
\begin{center}
\includegraphics[width=3.2in]{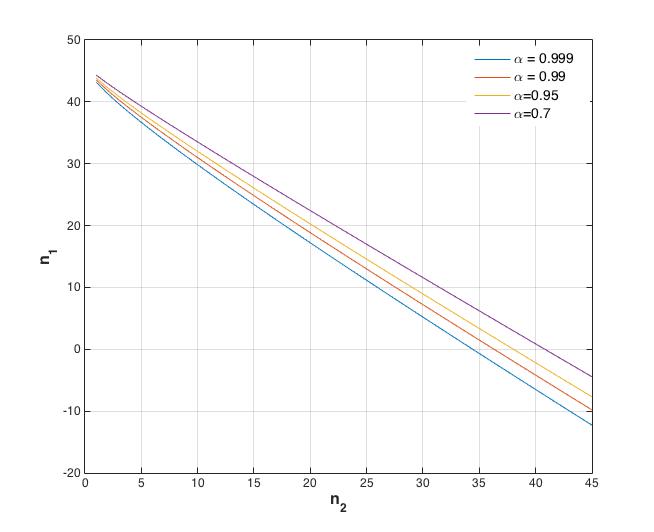}
\end{center}
\caption{Parameters: $\overline{A}=1$, $\underline{A}=0.3$, $\mu=0.65$, $V=30$.}\label{alpha:N:2classes}
\end{figure}

We next consider solving the offline problem when our goal is to schedule $N_1$ jobs of class 1 and $N_2$ jobs of class 2. The numbers $N_1$ and $N_2$ are given as an input, and our goal is to minimize the number of machines denoted by $M^*$, such that each machine is assigned a pair $(n_1, n_2)$ that satisfies equation \eqref{N:alpha:2}. Since $n_1$ and $n_2$ should be integer numbers, one can compute for a given value of $\alpha$, all the feasible pairs that lie on the curve or just below  (we have at most $K = \min_{i=1, 2} \max {n_i}$ such pairs). In other words, for each $k=1, 2, \ldots, K$, we compute a pair of coefficients denoted by $(\beta_k, \gamma_k)$. The optimization problem becomes a cutting stock problem:
\begin{equation}
\label{equation:cuttingstock:1D}
  \begin{array}{ll}
       M^* =  \displaystyle \min_{z_k}
      & \displaystyle \sum_{k=1}^K z_k
    \\[12pt]
    \text{s.t.}
      & \displaystyle \sum_{k=1}^K \beta_k z_k \geq N_1 
    \\[11pt]
      & \displaystyle \sum_{k=1}^K \gamma_k z_k  \geq N_2 
    \\[11pt]
      & z_k \geq 0, \text{integer} \qquad \forall k.
  \end{array}
\end{equation}
The decision variable $z_k$ represents the number of times we use the pair $(\beta_k, \gamma_k)$, and $M^*$ denotes the optimal number of machines. As a result, assuming that we have only two classes of jobs (with different $\mu$s and $b$s), one can solve the deterministic linear integer program in \eqref{equation:cuttingstock:1D} and obtain a solution for problem (SMBP). Note that the above treatment can easily be extended to more than two classes of jobs.

\section{Proof of Theorem \ref{firstfit94}}\label{proof94}

\proof{Proof.}
Let $n_1$ be the number of machines purchased by $\firstfit$ with only a single job, and $S_1$ be the set of $n_1$ jobs assigned to these machines. Similarly, we define $n_2$ to be the number of machines with at least two jobs, and $S_2$ be the set of their jobs. The goal is to prove that $n_1+n_2 \leq \frac{9}{4}\opt+1$. We know that any pair of jobs among the $n_1$ jobs in $S_1$ does not fit in a single machine (by the definition of $\firstfit$). Therefore, any feasible allocation (including the optimal allocation) needs at least $n_1$ machines. In other words, we have $\opt \geq n_1$. This observation also implies that the sum of $\mu_j+b_j$ for any pair of jobs in $S_1$ is greater than $\frac{3}{4}$ (using Lemma \ref{lem:lowerbound:34}). 
If we sum up all these inequalities for the different pairs of jobs in $S_1$, we have: $(n_1-1) \sum_{j \in {S_1}} (\mu_j + b_j) > {n_1 \choose 2} \frac{3}{4}$. We note that the $n_1-1$ term on the left side appears because every job $j \in S_1$ is paired with $n_1-1$ other jobs in $S_1$, and the ${n_1 \choose 2}$ term on the right side represents the total number of pairs of jobs in $S_1$. By dividing both sides of this inequality by $n_1-1$, we obtain $\sum_{j \in S_1} (\mu_j + b_j) > \frac{3n_1}{8}$.

We also lower bound $\sum_{j \in S_2} (\mu_j+b_j)$ as a function of $n_2$ as follows. Let $m_1 < m_2 < \cdots < m_{n_2}$ be the machines that have at least two jobs, and the ordering shows in which order they were purchased (e.g., $m_1$ was purchased first). Define $M_i$ to be the set of jobs in machine $m_i$. 
By definition of $\firstfit$, any job $j$ in machine $m_{i+1}$ could not be assigned to machine $m_i$ because of the feasibility constraints for any $1 \leq i < n_2$. In other words, the set of jobs $M_i$ with any job $j \in M_{i+1}$ form an infeasible set. Therefore, we have $\mu_j+b_{j} + \sum_{j' \in M_i} (\mu_{j'} + b_{j'}) > \frac{3}{4}$. 
For each $1 \leq i < n_2$, one can pick two distinct jobs $j_1$ and $j_2$ from $M_{i+1}$, and write the following two inequalities: 
$$
\mu_{j_1}+b_{j_1} + \sum_{j' \in M_i} (\mu_{j'} + b_{j'}) > \frac{3}{4}
~~~~~ \text{and}
~~~~
\mu_{j_2}+b_{j_2} + \sum_{j' \in M_i} (\mu_{j'} + b_{j'}) > \frac{3}{4}.
$$
Summing up these two inequalities implies that:
$
\mu_{j_1}+b_{j_1} + \mu_{j_2} + b_{j_2} + 2\sum_{j' \in M_i} (\mu_{j'} + b_{j'}) > \frac{3}{2}.
$
Since $j_1$ and $j_2$ are two distinct jobs in $M_{i+1}$, we have:
$$
\sum_{j \in M_{i+1}} (\mu_j + b_{j}) + 2\sum_{j' \in M_i} (\mu_{j'} + b_{j'}) > \frac{3}{2}.
$$  
Now we sum up this inequality for different values of $i \in \{1, 2, \cdots, n_2-1\}$ to achieve that:
$$
2 \sum_{j \in M_1} (\mu_j + b_{j}) + 3\sum_{i=2}^{n_2-1}\sum_{j \in M_i} (\mu_j + b_{j}) + \sum_{j \in M_{n_2}} (\mu_j + b_j) > \frac{3}{2} \times (n_2-1),
$$
and consequently, we have $3\sum_{i=1}^{n_2}\sum_{j \in M_i} (\mu_j + b_j) > \frac{3}{2} \times (n_2-1)$. This is equivalent to $\sum_{j \in S_2} (\mu_j + b_j) > \frac{1}{2} \times (n_2-1)$. Combining both inequalities, we obtain: $\sum_{j \in S_1 \cup S_2} (\mu_j+b_j) > \frac{3}{8}n_1 + \frac{1}{2}(n_2-1)$. On the other hand, $\opt$ is at least $\sum_{j \in S_1 \cup S_2} (\mu_j + b_j)$. We then have the following two inequalities: 
\begin{eqnarray}
\opt &\geq& n_1, \label{firstineq} \\
\opt &>& \frac{3}{8}n_1 + \frac{1}{2}(n_2-1). \label{secondineq}
\end{eqnarray} 
We can now multiply \eqref{firstineq} by $\frac{1}{4}$ and \eqref{secondineq} by $2$, and sum them up. We conclude that $\frac{9}{4}\opt + 1$ is greater than $n_1 + n_2$, which is the number of machines purchased by Algorithm $\firstfit$. $\Box$ 
\endproof

\section{Proof of Theorem~\ref{thm:full-homogeneous}}\label{proofopt}

We first state the following Lemma that provides a lower bound on $\opt$. For any $a, b \geq 0$, we define the function $f(a,b) = \frac{2a+b+\sqrt{b(4a+b)}}{2}$.

\begin{lemma}\label{lem:f-lowerbound}
For any feasible set of jobs $S$, the sum $\sum_{j \in S} f(\mu_j, b_j)$ is at most $1$. 
\end{lemma}

\proof{Proof.}
Define $\bar{\mu} = \frac{\sum_{j \in S} \mu_j}{|S|}$ and $\bar{b} = \frac{\sum_{j \in S} b_j}{|S|}$.
Since the function $f$ is concave with respect to both $a$ and $b$, using Jensen's inequality we have $\sum_{j \in S} f(\mu_j, b_j) \leq |S|f(\bar{\mu}, \bar{b})$. Since $S$ is a feasible set, $\cost(S) = |S|\bar{\mu} + \sqrt{|S|\bar{b}} \leq 1$. The latter is a quadratic inequality with variable $x = \sqrt{|S|}$, so that we we can derive an upper bound on $|S|$ in terms of $\bar{\mu}$ and $\bar{b}$. Solving the quadratic form $\bar{\mu}X + \sqrt{\bar{b}X} = 1$ yields:
$$
X = \frac{-\sqrt{\bar{b}} \pm \sqrt{\bar{b} + 4\bar{\mu}}}{2\bar{\mu}}.
$$
We then have $\sqrt{|S|} \leq \frac{\sqrt{4\bar{\mu}+\bar{b}} - \sqrt{\bar{b}}}{2\bar{\mu}} = \frac{2}{\sqrt{\left(4\bar{\mu}+\bar{b}\right)} + \sqrt{\bar{b}}}$.  Therefore, $|S| \leq \frac{4}{4\bar{\mu}+\bar{b} + \bar{b} + 2\sqrt{\bar{b}\left(4\bar{\mu}+\bar{b}\right)}} = \frac{1}{f(\bar{\mu}, \bar{b})}$, where the equality follows by the definition of the function $f$. Equivalently, $|S|f(\bar{\mu},\bar{b}) \leq 1$, and hence  $\sum_{j \in S} f(\mu_j, b_j) \leq 1$. $\Box$ 
\endproof

\proof{Proof of Theorem~\ref{thm:full-homogeneous}.}
For any arbitrary allocation of jobs, applying Lemma~\ref{lem:f-lowerbound} to all the machines implies that the number of machines is at least $\sum_{j=1}^N f(\mu_j, b_j)$. 
So it suffices to upper bound the number of machines as a function of $\sum_{j=1}^N f(\mu_j, b_j)$.
For any given machine, we prove that the sum of $f(\mu_j, b_j)$ for all jobs $j$ assigned to this machine is at least $1-O(\epsilon+\delta)$. 
Consider the set of jobs $S$ assigned to a given machine.
Similar to the proof of Lemma~\ref{lem:f-lowerbound}, we define  $\bar{\mu} = \frac{\sum_{j \in S} \mu_j}{|S|}$ and $\bar{b} = \frac{\sum_{j \in S} b_j}{|S|}$. 
Define $r = \frac{\sum_{j \in S} b_j}{\sum_{j \in S} \mu_j} = \frac{\bar{b}}{\bar{\mu}}$. 
We start by lower bounding $f(\mu_j, b_j)$ as a function of $f(\mu_j, r\mu_j)$. Recall that each machine is $\delta$-homogeneous, i.e., for all pairs of jobs in the same machine, the ratios $\frac{b_j}{\mu_j}$ are at most a multiplicative factor of $1+\delta$ away from each other. Hence, any ratio $\frac{b_j}{\mu_j}$ is at least $\frac{r}{(1+\delta)}$. Consequently, we have $b_j \geq \frac{r \mu_j} {1+\delta}$ which implies that:
$$
f(\mu_j, b_j) \geq f \big(\frac{\mu_j} {1 + \delta}, b_j \big) \geq f\big(\frac{\mu_j} {1 + \delta}, \frac{r \mu_j} {1 + \delta}\big) =  \frac{1} {1 + \delta} f(\mu_j, r \mu_j) \geq (1-\delta)f(\mu_j, r \mu_j).
$$
The first and second inequalities follow from the monotonicity of the function $f$, the equality follows from the definition of $f$, and the last inequality holds since $\delta \geq 0$. It now suffices to lower bound $\sum_{j \in S} f(\mu_j, r\mu_j)$ so as to obtain a lower bound for $\sum_{j \in S} f(\mu_j, b_j)$. 
We note that for any $\eta \geq 0$, we have $f(\eta \mu, \eta b) = \eta f(\mu,b)$ by the definition of $f$. 
Applying $\eta = \mu_j$, $\mu = 1$ and $b = r$, we obtain $f(\mu_j, r\mu_j) = \mu_j f(1, r)$ which implies:
$$
\sum_{j \in S} f(\mu_j, r\mu_j) = \sum_{j \in S} \mu_j f(1, r) = f(A,B),
$$
where $A = \sum_{j \in S} \mu_j$, and $B = r \sum_{j \in S} \mu_j = \sum_{j \in S} b_j$. The last equality above follows from $f(\eta \mu, \eta b) = \eta f(\mu,b)$ with $\eta = A$. Recall that each machine is  $\epsilon$-full, i.e., $\cost(S) \geq 1-\epsilon$ or equivalently, $A+\sqrt{B} \geq 1-\epsilon$. Let $A' = \frac{A}{(1-\epsilon)^2}$ and $B'= \frac{B}{(1-\epsilon)^2}$. Then, we have:
$$
A'+ \sqrt{B'} = \frac{A}{(1-\epsilon)^2} + \frac{\sqrt B}{1-\epsilon} \geq \frac{A+\sqrt{B}}{1-\epsilon} \geq 1.
$$
Since $B = rA$, we also have $B' = rA'$, and the lower bound can be rewritten as follows: 
$
A'+\sqrt{rA'} \geq 1,
$
which is the same quadratic form as in the proof of Lemma~\ref{lem:f-lowerbound}. Similarly, we prove that $A' \geq \frac{4}{4+2r+2\sqrt{r(4+r)}}$ which is equal to $\frac{1}{f(1,r)}$. We also know that $f(A',B') = f(A', rA') = A' f(1,r)$. We already proved that $A \geq \frac{1}{f(1,r)}$, so we have $f(A',B') \geq 1$. By definition of $A'$ and $B'$, we know that$f(A,B) = (1-\epsilon)^2f(A',B') \geq (1-\epsilon)^2$. We conclude that the sum $\sum_{j \in S} f(\mu_j, b_j) \geq (1-\delta) f(A,B) \geq (1-\delta)(1-\epsilon)^2$. As a result, for each machine the sum of $f(\mu_j,b_j)$ is at least $(1-\delta)(1-\epsilon)^2$. Let $m$ be the number of purchased machines. Therefore, the sum $\sum_{j=1}^N f(\mu_j,b_j)$ is lower bounded by $(1-\delta)(1-\epsilon)^2m$, and at the same time upper bounded by $\opt$. Consequently, $m$ does not exceed $\frac{\opt}{(1-\delta)(1-\epsilon)^2}$ and this concludes the proof.
$\Box$ 
\endproof

\section{Proof of Theorem \ref{offline2}}\label{proofoffline}

\proof{Proof.}
We first prove that the algorithm terminates in a finite number of iterations. Note that all the update operations (except the first one) reduce the number of purchased machines, and hence, there are no more than $N$ of those. As a result, it suffices to upper bound the number of times we perform the first update operation. Since we assign jobs to lower id machines, there cannot be more than $N^2$ consecutive first update operations. Consequently, after at most $N \times N^2 = N^3$ operations, the algorithm has to terminate. 

Next, we derive an upper bound on the number of purchased machines at the end of the algorithm. Note that all the machines belong to one of the following four categories:
\begin{itemize}
\item Single job machines, i.e., the set $A_1$.
\item Medium machines with only one non-good job -- denoted by the set $B$.
\item Medium machines with at least two non-good jobs -- denoted by the set $C$.
\item Large machines, i.e., the set $A_5$.
\end{itemize}
Let $a, b, c$, and $d$ be the number of machines in $A_1, B, C$, and $A_5$ respectively. 
Since no update operation is possible (as the algorithm already terminated), the $b$ non-good jobs assigned to the machines in the set $B$ do not fit in any of the single job machines, and no pair of them fit together in a new machine. Consider these $b$ non-good jobs in addition to the $a$ jobs in the machines of the set $A_1$. No pair of these $a+b$ jobs fit in one machine together and therefore, $\opt \geq a+b$. 

We also know that $\opt \geq \sum_{j=1}^{N} (\mu_j+b_j)$. Next, we derive a more elaborate lower bound on $\opt$ by writing the sum of $\mu_j+b_j$ as a linear combination of the sizes of the sets $A_1, B, C,$ and $A_5$. 
For each machine $i$, let $M_i$ be the set of jobs assigned to this machine.
Let $i_1 < i_2< \cdots < i_{d}$ be the indices of machines in the set $A_5$, where $d = |A_5|$. Since we cannot perform the first update operation anymore, we can say that no job in machine $i_{\ell+1}$ fits in machine $i_{\ell}$ for any $1 \leq \ell < d$. Therefore, $\mu_j+b_j + \sum_{j' \in M_{i_{\ell}}} (\mu_{j'} + b_{j'}) > \frac{3}{4}$ for any $j \in M_{i_{\ell+1}}$ (using Lemma \ref{lem:lowerbound:34}). We write this inequality for $5$ different jobs (arbitrarily chosen) in $M_{i_{\ell+1}}$ (recall that there are at least 5 jobs in this machine), and for all the values of $1 \leq \ell < d$. If we sum up all these $5 (d - 1)$ inequalities, then the right hand side would be $5 (d-1) \times \frac{3} {4}$. On the other hand, the term $\mu_j + b_j$ for every job in these $d$ machines appears on the left hand side at most $5+1=6$ times. Therefore, by summing up these inequalities, we obtain: $\sum_{i \in A_5} \sum_{j \in M_i} (\mu_j + b_j) > \frac{3}{4} \times \frac{5(d-1)}{6} = \frac{5(d-1)}{8}$. 

Each machine in $A_5$ has at least 5 jobs. Therefore, the term $\frac{5}{6}$ appears in the lower bound. With a similar argument, each machine in either $B$ or $C$ has at least 2 jobs and hence, this term is now replaced by $\frac{2}{3}$. The inequalities for every pair of machines in $B$ and $C$ are then: $\sum_{i \in B} \sum_{j \in M_i} (\mu_j+b_j) > \frac{3}{4} \times \frac{2(b-1)}{3} = \frac{b-1}{2}$, and $\sum_{i \in C} \sum_{j \in M_i} (\mu_j+b_j) > \frac{3}{4} \times \frac{2(c-1)}{3} = \frac{c-1}{2}$. 

Next, we lower bound $\sum_{i \in A_1 \cup C} \sum_{j \in M_i} (\mu_j+b_j)$ as a function of $a$ and $c$ in order to complete the proof. Recall that each machine in $C$ has at least two non-good jobs. If we pick one of these non-good jobs $j$, and a random machine $i$ from $A_1$, with probability at least $\frac{a-5}{a}$, job $j$ does not fit in machine $i$. This follows from the fact that a non-good job fits in at most 5 machines in $A_1$ and hence, a random machine in $A_1$ would not be be able to fit job $j$ with probability at least $\frac{a-5}{a}$. Therefore, $\mu_j+b_j + \sum_{j' \in M_i} (\mu_{j'} + b_{j'}) \geq \frac{3}{4}$ with probability at least $\frac{a-5}{a}$. We next consider two different cases depending on the value of $c$. 

If $c$ is at least $\frac{a}{2}$, we pick a random machine in $C$, and two of its non-good jobs $j_1$ and $j_2$ arbitrarily. We also pick a random machine in $A_1$. For each of these two jobs, the sum $\mu_j+b_j$ of the non-good job and the single job in the selected machine in $A_1$ is greater than $\frac{3}{4}$ with probability at least $\frac{a-5}{a}$. Summing up these two inequalities, we obtain:
\begin{align}\label{ineqthm6}
\frac{2}{a} \Big [ \sum_{i \in A_1} \sum_{j \in M_i} (\mu_j + b_j) \Big ] + \frac{1}{c} \Big [ \sum_{i \in C} \sum_{j \in M_i} (\mu_j + b_j) \Big ] > \frac{a-5}{a} \times \frac{3}{2}.
\end{align}
The left hand side of the above equation is composed of two terms. The first term is obtained through picking a random machine in $A_1$ (i.e., with probability $\frac{1}{a}$), and once the machine is picked, we sum up both equations so we obtain $\frac{2}{a}$. 
For the second term, every machine in the set $C$ is chosen with probability $\frac{1}{c}$.  When the machine is picked, we sum up on all the jobs and hence get an upper bound. 
As we have shown, $\sum_{i \in C} \sum_{j \in M_i} (\mu_j + b_j) \geq \frac{c-1}{2}$. 
Combining these two inequalities leads to (using $c \geq \frac{a}{2}$):
$$\sum_{i \in A_1 \cup C} \sum_{j \in M_i} (\mu_j + b_j) > \frac{a}{2} \times \frac{3(a-5)}{2a} + (1-\frac{a}{2c}) \times \frac{c-1}{2} \geq \frac{3a}{4} - \frac{15}{4} + \frac{c}{2} - \frac{a}{4} - \frac{1}{2} = \frac{a+c}{2} - 4.25.$$ 
By combining the three different bounds (on $A_1 \cup C$, $B$ and $A_5$), we obtain $\sum_{j=1}^N (\mu_j + b_j) \geq \frac{a+b+c}{2} + \frac{5d}{8} - 5.875$. Since $\opt \geq \sum_{j=1}^N (\mu_j + b_j)$, we conclude that the number of purchased machines $a+b+c+d$ is no more than $2\opt + 11$.

In the other case, we have $c < \frac{a}{2}$. Note that inequality \eqref{ineqthm6} still holds. However, since the coefficient $1 - \frac{a}{2c}$ becomes negative, we cannot combine the two inequalities as before. 
Instead, we lower bound the sum $\mu_j + b_j$ of jobs in $A_1$. We know that there is no pair of $a$ jobs in the $A_1$ machines that fit together in one machine. Therefore, $\sum_{i \in A_1} \sum_{j \in M_i} (\mu_j + b_j) \geq \frac{3a}{8}$. 
Next, we multiply inequality \eqref{ineqthm6} by $c$, and combine it with this new lower bound on  $\sum_{i \in A_1} \sum_{j \in M_i} (\mu_j + b_j)$, to obtain (using $c < \frac{a}{2}$):
$$\sum_{i \in A_1 \cup C} \sum_{j \in M_i} (\mu_j + b_j) > c \times \frac{3(a-5)}{2a} + (1-\frac{2c}{a}) \times \frac{3a}{8} > 
\frac{3c}{2} - \frac{5}{4} + \frac{3a}{8} - \frac{3c}{4} = 
\frac{3a}{8} + \frac{3c}{4} - \frac{5}{4}.
$$
Combining this inequality with similar ones on the sets $B$ and $A_5$, we obtain $\opt \geq \sum_{j=1}^N (\mu_j+b_j) > \frac{3a}{8} + \frac{b}{2} + \frac{3c}{4} + \frac{5d}{8} - \frac{19}{8}$. Finally, combining this with $\opt \geq a+b$ leads to $a+b+c+d \leq \frac{8}{5} \opt + \frac{2}{5} \opt + \frac{19}{5} = 2\opt + 3.75$, which concludes the proof. $\Box$ 
\endproof
\end{APPENDICES}
\end{document}